%% file: sample631.tex
\documentclass[twocolumn]{aastex631}

\usepackage{CJK}
\usepackage{bm} 
\usepackage{amsmath}
\usepackage{multirow}

\newcommand{\Secref}[1]{\hyperref[#1]{Section~\ref*{#1}}}
\newcommand{\Appref}[1]{\hyperref[#1]{Appendix~\ref*{#1}}}

\shorttitle{RV detection of exomoons}
\shortauthors{Ruffio et al.}

\graphicspath{{./}{figures/}}

\submitjournal{AJ}

\begin{document}
\begin{CJK*}{UTF8}{gbsn}

\title{Detecting exomoons from radial velocity measurements of self-luminous planets: application to observations of HR 7672 B and future prospects}

\newcommand{\petit}{\texttt{petitRADTRANS}}
\newcommand{\emcee}{\texttt{emcee}}
\newcommand{\pmn}{\texttt{PyMultiNest}}
\newcommand{\moog}{\texttt{MOOG}}
\newcommand{\teff}{T$_{\rm eff}$}
\newcommand{\feh}{\rm{[Fe/H]}}
\newcommand{\caltech}{Department of Astronomy, California Institute of Technology, Pasadena, CA 91125, USA}
\newcommand{\gps}{Division of Geological \& Planetary Sciences, California Institute of Technology, Pasadena, CA 91125, USA}
\newcommand{\ucsc}{Department of Astronomy \& Astrophysics, University of California, Santa Cruz, CA95064, USA}
\newcommand{\keck}{W. M. Keck Observatory, 65-1120 Mamalahoa Hwy, Kamuela, HI, USA}
\newcommand{\ucla}{Department of Physics \& Astronomy, 430 Portola Plaza, University of California, Los Angeles, CA 90095, USA}
\newcommand{\jpl}{Jet Propulsion Laboratory, California Institute of Technology, 4800 Oak Grove Dr.,Pasadena, CA 91109, USA}
\newcommand{\ucsd}{Center for Astrophysics and Space Sciences, University of California, San Diego, La Jolla, CA 92093}
\newcommand{\ifahonolulu}{Institute for Astronomy, University of Hawai`i, 2680 Woodlawn Drive, Honolulu, HI 96822, USA}
\newcommand{\berkeley}{Department of Astronomy, University of California at Berkeley, CA 94720, USA}
\newcommand{\ames}{NASA Ames Research Centre, MS 245-3, Moffett Field, CA 94035, USA}

\correspondingauthor{Jean-Baptiste Ruffio}
\email{jruffio@caltech.edu}

\author[0000-0003-2233-4821]{Jean-Baptiste Ruffio}
\affiliation{\caltech}
\affiliation{Center for Astrophysics and Space Science, University of California, San Diego; La Jolla, CA 92093, USA}

\author[0000-0001-9708-8667]{Katelyn Horstman}
\affiliation{\caltech}

\author[0000-0002-8895-4735]{Dimitri Mawet}
\affiliation{\caltech}
\affiliation{\jpl}

\author[0000-0001-8391-5182]{Lee J. Rosenthal}
\affiliation{\caltech}

\author[0000-0002-7094-7908]{Konstantin Batygin}
\affiliation{Division of Geological and Planetary Sciences California Institute of Technology, Pasadena, CA 91125, USA}

\author[0000-0003-0774-6502]{Jason J. Wang (王劲飞)}
\affiliation{Center for Interdisciplinary Exploration and Research in Astrophysics (CIERA) and Department of Physics and Astronomy, Northwestern University, Evanston, IL 60208, USA}

\author[0000-0001-6205-9233]{Maxwell Millar-Blanchaer}
\affiliation{Department of Physics, University of California, Santa Barbara, Santa Barbara, California, USA}

\author[0000-0002-4361-8885]{Ji Wang (王吉)}
\affiliation{Department of Astronomy, The Ohio State University, 100 W 18th Ave, Columbus, OH 43210 USA}

\author[0000-0003-3504-5316]{Benjamin J. Fulton}
\affiliation{NASA Exoplanet Science Institute, Caltech/IPAC-NExScI, 1200 East California Boulevard, Pasadena, CA 91125, USA}

\author[0000-0002-9936-6285]{Quinn M. Konopacky}
\affiliation{Center for Astrophysics and Space Science, University of California, San Diego; La Jolla, CA 92093, USA}

\author[0000-0003-2429-5811]{Shubh Agrawal}
\affiliation{\caltech}
\affiliation{Department of Physics and Astronomy, University of Pennsylvania, Philadelphia, PA 19104, USA}

\author[0000-0001-8058-7443]{Lea A. Hirsch}
\affiliation{Department of Chemical \& Physical Sciences, University of Toronto Mississauga, 3359 Mississauga Road Mississauga , Ontario L5L 1C6m, Canada}

\author[0000-0001-8638-0320]{Andrew W. Howard}
\affiliation{\caltech}

\author[0000-0002-3199-2888]{Sarah Blunt}
\affiliation{\caltech}
\altaffiliation{NSF Graduate Research Fellow}

\author[0000-0001-6975-9056]{Eric Nielsen}
\affiliation{Department of Astronomy, New Mexico State University, P.O. Box 30001, MSC 4500, Las Cruces, NM 88003, USA}


\author{Ashley Baker}
\affiliation{\caltech}

\author{Randall Bartos}
\affiliation{\jpl}

\author{Charlotte Z. Bond}
\affiliation{UK Astronomy Technology Centre, Royal Observatory, Edinburgh EH9 3HJ, United Kingdom}

\author{Benjamin Calvin}
\affiliation{\caltech}
\affiliation{\ucla}

\author{Sylvain Cetre}
\affiliation{\keck}

\author[0000-0001-8953-1008]{Jacques-Robert Delorme}
\affiliation{\keck}
\affiliation{\caltech}

\author{Greg Doppmann}
\affiliation{\keck}

\author[0000-0002-1583-2040]{Daniel Echeverri}
\affiliation{\caltech}

\author[0000-0002-1392-0768]{Luke Finnerty}
\affiliation{\ucla}

\author[0000-0002-0176-8973]{Michael P. Fitzgerald}
\affiliation{\ucla}

\author[0000-0001-5213-6207]{Nemanja Jovanovic}
\affiliation{\caltech}

\author{Ronald L\'opez}
\affiliation{\ucla}

\author[0000-0002-0618-5128]{Emily C. Martin}
\affiliation{\ucsc}

\author{Evan Morris}
\affiliation{\ucsc}

\author{Jacklyn Pezzato}
\affiliation{\caltech}

\author[0000-0003-4769-1665]{Garreth Ruane}
\affiliation{\caltech}
\affiliation{\jpl}

\author[0000-0003-1399-3593]{Ben Sappey}
\affiliation{\ucsd}

\author{Tobias Schofield}
\affiliation{\caltech}

\author{Andrew Skemer}
\affiliation{\ucsc}

\author{Taylor Venenciano}
\affiliation{Physics and Astronomy Department, Pomona College, 333 N. College Way, Claremont, CA 91711, USA}

\author{J. Kent Wallace}
\affiliation{\jpl}

\author[0000-0003-0354-0187]{Nicole L. Wallack}
\affiliation{Earth and Planets Laboratory, Carnegie Institution for Science, Washington, DC 20015, USA}

\author{Peter Wizinowich}
\affiliation{\keck}

\author[0000-0002-6618-1137]{Jerry W. Xuan}
\affiliation{\caltech}



\begin{abstract}
The detection of satellites around extrasolar planets, so called exomoons, remains a largely unexplored territory. In this work, we study the potential of detecting these elusive objects from radial velocity monitoring of self-luminous directly imaged planets. This technique is now possible thanks to the development of dedicated instruments combining the power of high-resolution spectroscopy and high-contrast imaging. First, we demonstrate a sensitivity to satellites with a mass ratio of $1-4\%$ at separations similar to the Galilean moons from observations of a brown-dwarf companion (HR 7672 B; K$_\text{mag}$=13; $0.7^{\prime\prime}$ separation) with the Keck Planet Imager and Characterizer (KPIC; $R\sim35,000$ in $K$ band) at the W. M. Keck Observatory. Current instrumentation is therefore already sensitive to large unresolved satellites that could be forming from gravitational instability akin to binary star formation. Using end-to-end simulations, we then estimate that future instruments such as MODHIS, planned for the Thirty Meter Telescope, should be sensitive to satellites with mass ratios of $\sim10^{-4}$.  Such small moons would likely form in a circumplanetary disk similar to the Jovian satellites in the solar system. Looking for the Rossiter–McLaughlin effect could also be an interesting pathway to detecting the smallest moons on short orbital periods. Future exomoon discoveries will allow precise mass measurements of the substellar companions that they orbit and provide key insight into the formation of exoplanets. They would also help constrain the population of habitable Earth-sized moons orbiting gas giants in the habitable zone of their stars.
\end{abstract}



\input{content}


\begin{acknowledgments}
J.-B. R. acknowledges support from the David and Ellen Lee Prize Postdoctoral Fellowship.

Funding for KPIC has been provided by the California Institute of Technology, the Jet Propulsion Laboratory, the Heising-Simons Foundation through grants \#2019-1312 and \#2015-129, the Simons Foundation, and the United States National Science Foundation Grant No. AST-1611623.

Ji Wang acknowledges the support by the National Science Foundation under Grant No. 2143400.

The W. M. Keck Observatory is operated as a scientific partnership among the California Institute of Technology, the University of California, and NASA. The Keck Observatory was made possible by the generous financial support of the W. M. Keck Foundation. We also wish to recognize the very important cultural role and reverence that the summit of Maunakea has always had within the indigenous Hawaiian community. We are most fortunate to have the opportunity to conduct observations from this mountain.
\end{acknowledgments}

%

\vspace{5mm}
\facilities{Keck II (KPIC)}


\software{astropy\footnote{\url{http://www.astropy.org}} \citep{Astropy2013}, Matplotlib\footnote{\url{https://matplotlib.org}} \citep{Hunter2007},
PSIsim\footnote{\url{https://github.com/planetarysystemsimager/psisim}},
RVSearch\footnote{\url{https://github.com/California-Planet-Search/rvsearch}} \citep{Rosenthal2021},
KPIC Data Reduction Pipeline\footnote{\url{https://github.com/kpicteam/kpic_pipeline}} \citep{Delorme2021b},
BREADS\footnote{\url{https://github.com/jruffio/breads}} \citep{Ruffio2021, Agrawal2022},
}



\appendix
\input{appendix}


\bibliography{bibliography}{}
\bibliographystyle{aasjournal}



\end{CJK*}
\end{document}

%% file: content.tex
\section{Introduction}
\label{sec:introduction}

\subsection{Exomoon formation pathways}

Moons similar to those around Jupiter are expected to form in circumplanetary disks (CPD) as a by-product of planet formation \citep{Batygin2020}. The typical CPD total dust mass relative to the planet is around $10^{-4}$ \citep{Canup2006} commensurate with the mass ratios of solar system satellites around the gas giants listed in \autoref{tab:solarsystemmoons}. This is consistent with the measured value of the CPD around PDS 70 c from ALMA observations \citep{Benisty2021}, which is $0.031M_{\mathrm{Earth}}$ and corresponds to a mass ratio of about $5\times10^{-5}$, assuming a $2M_{\mathrm{Jup}}$ planet \citep{Benisty2021}. It is also possible to form larger moons from the merger of Galilean-like multiple systems. This is the proposed scenario to explain the high-eccentricity and large mass of Saturn's moon Titan \citep{Asphaug2018}. Alternative formation pathways include the capture of satellites (e.g., Neptune's moon Triton, \citet{Agnor2006}), collisions with protoplanets (e.g., the Moon, \citet{Canup2001}]), or even gravitational instability like in the formation of brown dwarf binaries \citep{Lazzoni2020}. The detection of satellites that formed in a CPD remains challenging with current instrumentation, but binary planets and brown dwarfs are already accessible with various techniques depending on their separation \citep{Lazzoni2022}. Characterizing the different populations with different mass ratios, such as binary brown-dwarf companions (ie, triple systems) and smaller CPD moons, could help inform the formation pathways of directly imaged planets. Indeed, binary companions could only occur in a top-down scenario, while CPD formation could occur in all cases. An important lesson from early exoplanet discoveries is that the solar system is not a good predictor of exoplanet demographics. Planet formation theories also often struggle to account for the diversity of new discoveries. For example, the first discoveries of hot Jupiters and the ubiquity of super-Earths and mini Neptunes were initially a surprise to the community \citep{Batalha2014}. As a corollary, it would be unwise to assume that exomoon searches should be any different \citep{Kipping2015}. It is therefore important that we keep pushing the discovery space with new observational methods.

\subsection{Status of exomoon searches}

For the past decade, transiting surveys have unequivocally dominated the landscape of exomoon searches \citep{Kipping2012} through the analysis of transit timing variations and additional transit signal from the moon. They have placed the first constraints on exomoon occurrence rates and shown that high-mass ratio satellites are not common around short-period exoplanets \citep{Kipping2015,Teachey2018a}. Other detection techniques have been used to look for exomoons around directly imaged exoplanets such as astrometry or direct imaging. While these terms also refer to planet detection techniques, in this context, direct imaging means to spatially resolve the satellite from the planet. Astrometric detections refers to the measurement of the astrometric wobble of a planet caused by orbiting moons with precise interferometric instruments such as VLTI/GRAVITY \citep{Abuter2021}.
To date, only a handful of exomoon candidates have been proposed: for example two around transiting planets \citep{Teachey2018b,Kipping2022}, one orbiting a directly-imaged brown dwarf \citep{Lazzoni2020}, and another around an isolated planetary mass object \citep{Limbach2021}. None have been confirmed. Most notably, the exomoon candidate Kepler 1708 b-i is a transiting 2.6 Earth radii object orbiting a Jupiter-sized planet, which, if confirmed, would be several orders of magnitude larger than the Galilean moons in terms of mass ratio \citep{Kipping2022}.

Transiting planets generally have short orbital periods and smaller Hill spheres. These conditions could be less favorable to moon formation and retention, while observed transits of long period exoplanets ($>1\,\mathrm{au}$) are intrinsically rare. For example, it has been suggested that planets could lose their satellites as they migrate inward \citep{Spalding2016}. The detection of exomoons around imaged planets with astrometry or direct imaging is more sensitive to longer-period moons \citep{Lazzoni2022}.
These two techniques might not be well suited for satellites with mass ratios and separations ($<30R_{\mathrm{Jup}}$) similar to the ones orbiting the solar system gas giants.

\subsection{A promising alternative: RV detections around directly imaged planets}

Another technique has been proposed to look for moons around directly imaged planets using RV measurements of the planet itself \citep{Vanderburg2018,Vanderburg2021}.
By measuring the wobble of planets caused by orbiting satellites, planetary RV surveys is a promising alternative for finding Galilean moons analogs.
Directly-imaged companions are likely to have a different formation and migration history compared to transiting exoplanets. They are also generally more massive and further away resulting in much more extended Hill spheres. Models suggest that larger planets form even larger moons following the scaling $m\propto M^{3/2}$, with m and M the masses of the moon and the planet respectively (based on equation 43 in \citet{Batygin2020}).
As another formation pathway, binary systems forming as the tail end of stellar formation through gravitational instability of the protostellar cloud or the protoplanetary disk could lead to a population of easily detectable high-mass ratio satellites and binary companions \citep{Lazzoni2022}. 
In summary, directly imaged companions could be more likely to host larger moons, which could be detected from RV monitoring of the planets themselves.

\subsection{Recent technology developments}

There are two aspects to optimizing the choice of a target for exomoon searches: the RV precision that can be achieved and the probability of the object to host a moon. In this work, we focus on the former and study the detectable mass ratios for exomoons as a function of the instrument, the telescope, and the planet or brown-dwarf properties.
Although the RV detection of exomoons remains challenging due to the intrinsic faintness of planets and the light contamination from the glare of the host star, the expertise obtained from 30 years of stellar RV exoplanet detections is an invaluable asset. Indeed, stable high-resolution spectrographs and data analysis techniques are already demonstrating stability and performance in excess of the level of precision needed for exomoons.  \citet{Vanderburg2021} derived the first exomoon mass upper limits with this technique around the HR 8799 planets based on the planetary RV time series reported by \citet{Ruffio2021} using OSIRIS, an $R=4,000$ spectrograph, at the W. M. Keck observatory. This study ruled out moons with mass greater than a Jupiter mass and period less than one day that would be orbiting the 7 Jupiter mass planet HR 8799 c. 

Recent technological advances in infrared high-resolution spectroscopy for high-contrast companions are enabling the first planetary RV searches for exomoons \citep{Snellen2015,Jovanovic2017,Delorme2021b,Otten2021}.
The Keck Planet Imager and Characterizer (KPIC) recently demonstrated $R=35,000$ $K$-band spectroscopy of directly-imaged exoplanets, including the HR 8799 system, measuring their RVs and obtaining spin measurements for the first time \citep{Wang2021}. KPIC is the first implementation of a new class of spectrographs that combines the power of the Keck II adaptive optics systems, the stability and starlight suppression of single mode fibers, and the high spectral resolution of the NIRSPEC spectrograph for the detailed study of directly imaged planets \citep{Delorme2021b}. 
We observed a bright brown dwarf companion (HR 7672 B, $K=13.04$, $\sim 0.7^{\prime\prime}$, \citet{Liu2002,Boccaletti2003}) as part of the commissioning and science verification of KPIC \citep{Delorme2021b, WangJi2022}.
While it is at the boundary of the stellar regime, HR 7672 B is an interesting benchmark companion, because it has a dynamically measured mass of $72.7\pm0.8 M_{\rm{Jup}}$ \citep{Crepp2012,Brandt2019} and its composition should be similar to that of the star due to its assumed formation history from gravitational instability. Using this target, \citet{WangJi2022} showed that accurate atmospheric compositions could be retrieved using KPIC's high resolving power and angular resolution by demonstrating a $1.5\sigma$ consistency between the composition of HR 7672 B and its host star (see also \citealt{Xuan2022}). 
HR 7672 B was also observed for a full night with KPIC as a test case for variability studies. This time series can be used to put the deepest limits to date on the mass of an orbiting satellite around the sub-stellar companion, which we are demonstrating in this work.
KPIC is already undergoing several upgrades including a laser frequency comb which will enable precise RV science \citep{Yi2016,Jovanovic2020}. The expected doubling of the instrumental throughput will significantly improve its sensitivity \citep{Jovanovic2020,Echeverri2022}. The next generation of high-contrast high-resolution spectrograph such as HISPEC at the W. M. Keck Observatory and MODHIS on the future Thirty Meter Telescope (TMT) will undoubtedly open new frontiers in this field by allowing $0.98-2.46\,\mu\mathrm{m}$ simultaneous coverage at an average spectral resolution $R>100,000$ \citep{Mawet2019,Mawet2022}.

\subsection{Outline}
In \autoref{sec:HR7672B}, we present exomoon RV detection limits for the brown dwarf companion HR 7672 B using KPIC. In \autoref{sec:futurefacilities}, we then simulate observations of the same brown dwarf and the planet HR 8799 c with next generation facilities and compare their sensitivity to the moons in the solar system. 
In \autoref{sec:generalcase}, we explore the parameter space of satellites that could be detected with TMT/MODHIS as a function of planet properties. Finally, we conclude on the prospects for RV detections of exomoons in \autoref{sec:discussion}.

\begin{table*}[ht]
  \centering
\begin{tabular}{ cccccccc } 
 \hline
  Moon & Planet & Mass & Mass ratio & Semi-Major axis &  Period & RV semi-amplitude \\
   &  & ($M_{\Earth}$) &  & ($R_{\mathrm{Jup}}$) &  (day) &  ($\mathrm{m}/\mathrm{s}$) \\ 
 \hline
Io & Jupiter &  $1.50\times10^{-2}$ & $4.71\times10^{-5}$ & 5.90 & 1.77 & 0.82 \\
Europa & Jupiter & $8.04\times10^{-3}$ & $2.53\times10^{-5}$ & 9.39 & 3.55 & 0.35 \\
Ganymede & Jupiter & $2.48\times10^{-2}$ & $7.81\times10^{-5}$ & 14.97 & 7.15 & 0.85 \\
Callisto & Jupiter & $1.80\times10^{-2}$ & $5.67\times10^{-5}$ & 26.33 & 16.69 & 0.46 \\
Titan & Saturn & $2.25\times10^{-2}$ & $2.37\times10^{-4}$ & 17.09 & 15.95 & 1.32 \\
Titania & Uranus &  $5.73\times10^{-4}$ & $3.94\times10^{-5}$ & 6.10 & 8.71 & 0.14 \\
Oberon & Uranus &  $4.82\times10^{-4}$ & $3.32\times10^{-5}$ & 8.16 & 13.46 & 0.10 \\
Triton & Neptune &  $3.58\times10^{-2}$ & $2.09\times10^{-4}$ & 4.96 & -5.88 & 0.92 \\
\hline
Kepler-1708 b-i & Kepler-1708 b & $<37$ & $<0.11 (2\sigma)$ &  & & & \\
\hline
\end{tabular}
\caption{Properties of the largest satellites orbiting the solar system gas giants from the NASA Space Science Data Coordinated Archive (\url{https://nssdc.gsfc.nasa.gov/planetary/}). The negative period of Triton is indicating its retrograde orbit. Kepler-1708 b-i is a transiting exomoon candidate \citep{Kipping2022}. The period and RV semi-amplitude for these moons can also be found in \citet{Vanderburg2018}.}

\label{tab:solarsystemmoons}
\end{table*}

\section{Exomoon limits around HR 7672 B with KPIC}
\label{sec:HR7672B}

\subsection{Observations and data reduction}
The brown dwarf companion HR 7672 B was observed three times in 2020 and then for a full night on July 4, 2021 with KPIC ($R\sim35,000$) in $K$ band ($1.9-2.4\,\mu$m) \citep{Mawet2017,Delorme2021b}. These observations are detailed in \autoref{tab:observations}. The first three epochs included one to two hours of on-target exposures per night and were already published in \citet{WangJi2022} and \citet{Delorme2021b}.
Unfortunately, the conditions on July 4, 2021 were well below average with the companion undetectable in some individual 5-minute exposures. During this one night specifically, we used an ABAB pattern to nod the companion between two KPIC fibers, fiber 1 and 2, to limit or identify any fiber-specific biases. There was no nodding during the other epochs.
The data was reduced with the KPIC data reduction pipeline (DRP)\footnote{\url{https://github.com/kpicteam/kpic_pipeline}} following the same approach described in \citet{Wang2021,WangJi2022}.
The first steps include background subtraction, bad pixel correction, and the calibration of the fiber trace location and width on the detector for each NIRSPEC spectroscopic order. Optimal extraction is then used to extract the spectra and the wavelength solution is derived from the telluric and stellar lines of a M giant, namely HIP 81497, taken on the same night. For this purpose, the telluric model is generated with the Planetary Spectrum Generator \citep{Villanueva2018} and star is modeled by a Phoenix model ($\log(g/[1\,\mathrm{cm.s}^{-2}])=1$; $T_{\mathrm{eff}}=3600\,\mathrm{K}$ \citet{Husser2013}).

\begin{table*}
  \centering
\begin{tabular}{ cccccc } 
 \hline
  Object & Date & Exposure time & Seeing & Throughput \\ 
 \hline
HR 7672 B & 2020-06-08 & $11\times10$ min &$0.4^{\prime\prime}$& 1\% \\
HR 7672 B & 2020-06-09 & $10\times10$ min &$0.6^{\prime\prime}$& 1.5\% \\
HR 7672 B & 2020-09-28 & $7\times10$ min &$0.4^{\prime\prime}$ & 2.7\% \\
HR 7672 B & 2021-07-04 & $61\times5$ min & $1^{\prime\prime}$  & 2\% \\
 \hline
\end{tabular}
\caption{$K$-band observations of HR 7672 A and B with KPIC. The quoted throughput is the end-to-end from the top of the atmosphere, which is a better proxy of performance than the seeing for KPIC.}
\label{tab:observations}
\end{table*}

\subsection{Forward model and likelihood}
\label{sec:FM}
We use a forward modelling approach similar to \citet{Wang2021} and \citet{Ruffio2021} to measure the RV of HR 7672 B, which includes a joint modelling of the starlight and the companion signal. \citet{Wang2021spie} showed that the continuum could be included in the forward model with a fourth order polynomial, therefore not requiring the data to be high-pass filtered nor continuum normalized. In this work, we model the continuum using a spline-based linear model, which can be analytically marginalized using the general purpose python module \texttt{breads}\footnote{\url{https://github.com/jruffio/breads}} (Broad Repository for Exoplanet Analysis, Discovery, and Spectroscopy) based on the formalism in \citet{Ruffio2019}. The spline forward modeling has the advantage of being more robust to bad pixels than a Fourier based high-pass filter and avoids the non-linearity of a sliding-window median filter. The spline parameters are also easier to optimize than the coefficients of a high order polynomial for example.

\begin{figure*}
  \centering
  \includegraphics[trim={0cm 0cm 0cm 0cm},clip,width=1\linewidth]{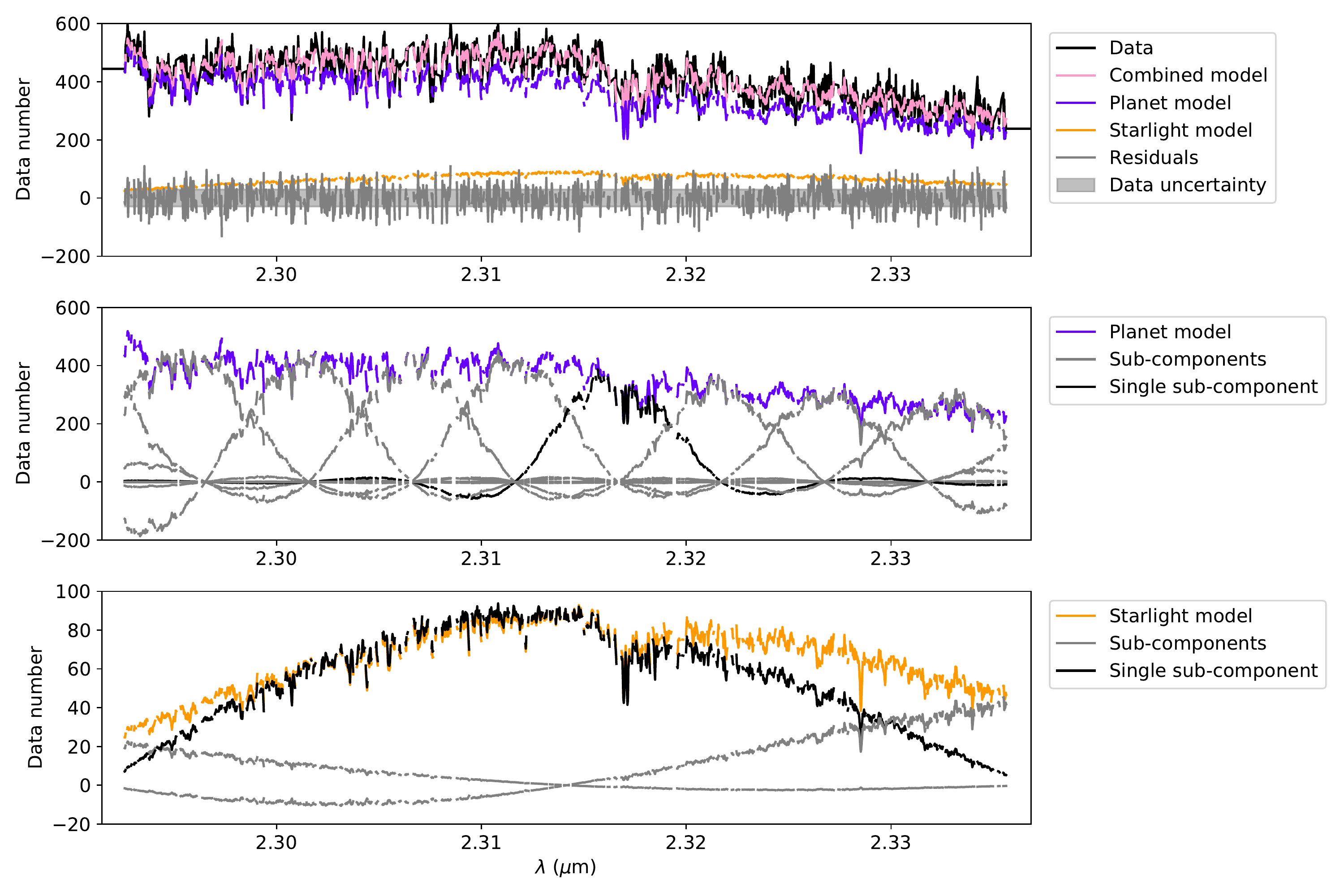}
  \caption{Illustration of the forward model used to derive the RV of HR 7672 B. This figure shows a single NIRSPEC order overlapping with the CO bandhead. (Top panel) A planet and a starlight model are jointly fitted to the data to account for the diffracted starlight contamination at the location of the companion. The data uncertainty measured by the KPIC DRP (shaded grey) slightly underestimates the amplitude of the residuals. (Center panel) The planet model is itself made of a linear combination of ten spline modes to model the continuum of the companion spectrum. (Bottom panel) The starlight intensity is also fitted with a spline using three nodes to account for speckles crossing at the location of the fiber. This flexible model of the continuum is an alternative to high-pass filtering and continuum normalization of high-resolution spectra.}
  \label{fig:fm}
\end{figure*}

We define the forward model as,
\begin{equation}
     {\bm d} = {\bm M}_{\mathrm{RV}}{\bm \phi} + {\bm n},
\end{equation}
where ${\bm d}$ is the data vector of size $N_d$, ${\bm M}_{\mathrm{RV}}$ is the linear model, ${\bm \phi}$ are the linear parameters, and ${\bm n}$ is a random vector of the noise with a diagonal covariance matrix ${\bm \Sigma}$. Off-diagonal elements in the covariance matrix are neglected here, but subsequent data processing steps would correct for this inaccuracy.  The different column vectors of the linear model are illustrated in \autoref{fig:fm}.
The data vector and the standard deviation of the noise used to define ${\bm \Sigma}_0$ are direct outputs of the KPIC data reduction pipeline. The variance of the noise is multiplied by a free parameter scaling factor $s^2$ to account for any underestimation of the noise.

KPIC includes four single mode fibers separated by $0.8^{\prime\prime}$ on a line. We can therefore acquire simultaneous spectra of the companion and the host star, more specifically the speckle field, by rotating the field of view using the Keck II adaptive optics system front-end K-mirror rotator. The observations of the star are used to derive simultaneous empirical models of the transmission and the starlight spectra used in the forward model. The starlight is used to model the speckle noise leaking into the fiber at the position of the companion. The wavelength calibration is different in each fiber so the spectra are linearly interpolated to match the sampling of the science fiber.
The planet model is defined as the spin-broadened best fit model from \citet{WangJi2022} using petitRADTRANS \citep{Molliere2019} multiplied by the empirical telluric and instrument transmission profile.
The continuum of both the planet and the speckle are modulated by a $3^{\rm rd}$ order spline model. Ten spline nodes are used in each spectral order ($\Delta \lambda \sim0.05\,\mu$m) for the planet model to manage any inaccuracies in the continuum due to imperfections in the atmosphere model fit. This number of nodes is analogous to a 200 pixel-wide high-pass filter. The number of nodes was chosen as a trade-off between the number of additional parameters and the optimal high-pass filter scale of 100 pixels found in \citet{Xuan2022}. The speckle continuum is modeled with three spline nodes to model any speckle crossing the fiber location as the wavelength changes.
This results in 13 linear parameters per spectral order representing the values of the continua at the location of the nodes (See \autoref{fig:fm}).
This defines the linear model ${\bm M}_{\mathrm{RV}}$ with dimensions $N_d\times13$, which is also a function of the RV of the planet, the only non-linear parameter fitted for here.

KPIC data features strong spectral fringing due to the Fabry–P\'{e}rot cavities formed by the transmissive optics inside the NIRSPEC spectrograph \citep{Hsu2021} and within the KPIC fiber injection unit \citep{Finnerty2022}. This effect is made worse by the high spatial coherence of the wavefront in KPIC. We therefore apply a Fourier filter to the data and the forward model by zeroing frequencies corresponding to the fringes. A physical model of the fringing such as \citet{Cale2019} could be explored in the future.

The likelihood function is defined from a multivariate Gaussian distribution as,
\begin{multline}
    \mathcal{L}(\mathrm{RV},{\bm \phi},s^2) =\frac{1}{\sqrt{(2\pi)^{N_d}\vert{\bm \Sigma}_0\vert s^{2N_d}}}\\  \exp\left[-\frac{1}{2s^2}({\bm d}-{\bm M}_{\mathrm{RV}}{\bm \phi})^{\top}{\bm \Sigma}_0^{-1}({\bm d}-{\bm M}_{\mathrm{RV}}{\bm \phi})\right].
\end{multline}
The likelihood is maximized using a linear least square solver on a grid of RV values from $-400$ to $400\,$km/s in steps of $0.2\,$km/s. The $1\sigma$ RV uncertainties are derived from the RV posterior calculated analytically according to Equation 10 in \citet{Ruffio2021} on this RV sampling. This method analytically marginalized the RV posterior for the modulation of the continuum and the noise scaling factor. The linear spline parameters used to fit the continuum are forced to be positive. This is theoretically inconsistent with the framework, which assumes unconstrained parameters, but it does not appear to significantly impact the RV time series.

Only the three reddest orders, out of nine in $K$ band, are used in this analysis. The bluest three orders (numbered 39-37; $1.94-2.09\,\mu$m) were discarded because they feature strong saturated CO$_2$ telluric lines that are generally harder to model, but also make for an unstable fit due to overlapping frequencies with the fringing and the simple Fourier filter. The middle three orders ($2.10-2.27\,\mu$m) lack sufficient stellar and telluric spectral lines to calibrate the wavelength precisely enough. Thus, only the remaining three orders are used in this analysis: $2.29-2.34\,\mu$m (order 33), $2.36-2.41\,\mu$m (order 32), and $2.44-2.49\,\mu$m  (order 31). Order 33 includes the carbon monoxide bandhead and therefore results in the strongest signal-to-noise ratio (S/N) and the most precise radial velocity measurement. Each NIRSPEC spectral order is fitted separately resulting in three RV estimates for each exposure.

\subsection{RV measurements}
\begin{figure*}
  \centering
  \includegraphics[trim={0cm 0cm 0cm 0cm},clip,width=1\linewidth]{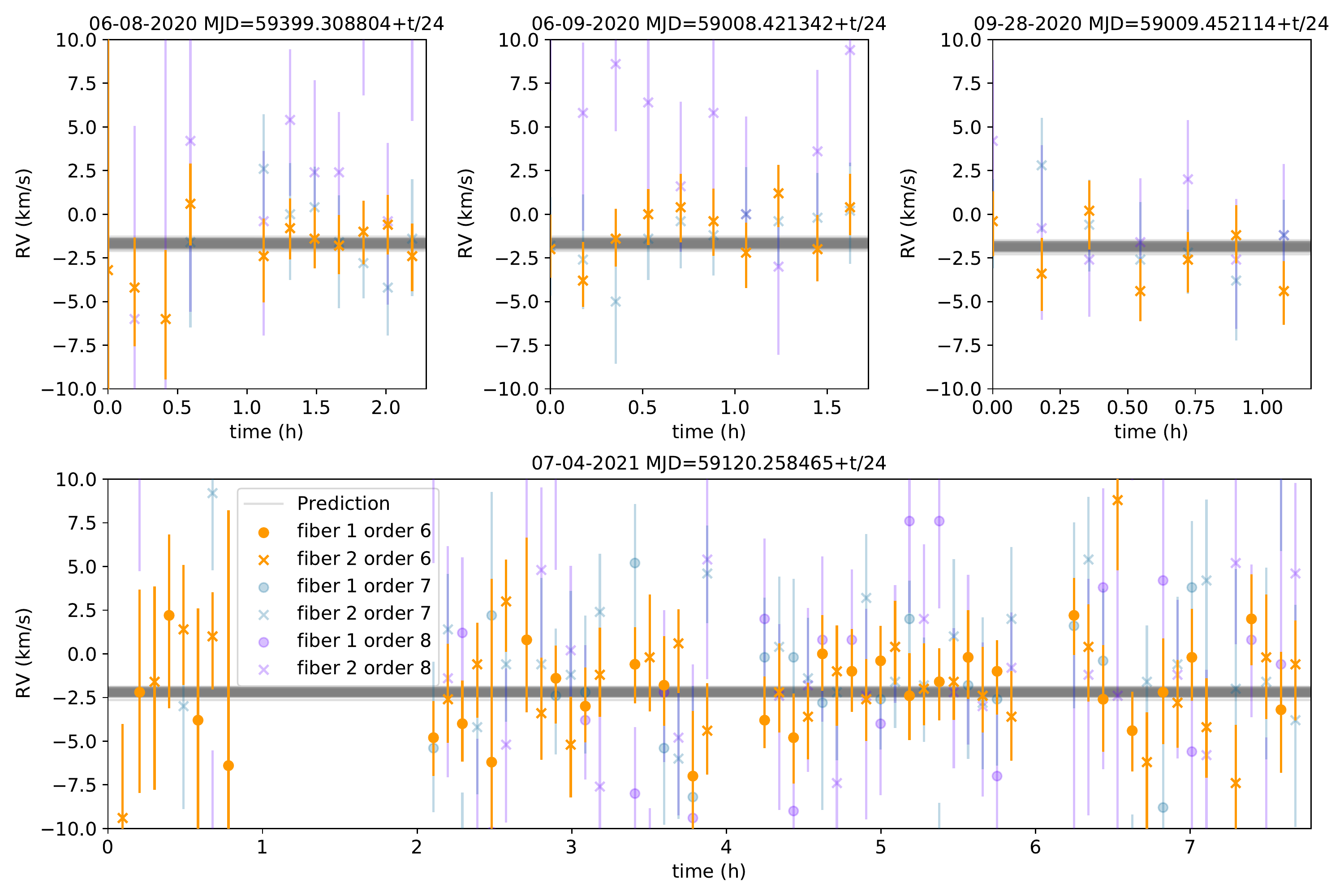}
  \caption{Measured RVs of HR 7672 B with KPIC. The grey lines are predicted RVs from one hundred posterior samples of the orbital motion of the brown dwarf.}
  \label{fig:rvs}
\end{figure*}

The barycentric corrected RV measurements for the four epochs and three orders are shown in \autoref{fig:rvs}. 
Following the method described in \autoref{sec:FM}, the median RV uncertainties in five minute exposures are $2.5\,\mathrm{km.s^{-1}}$,$4.0\,\mathrm{km.s^{-1}}$,$4.9\,\mathrm{km.s^{-1}}$ for order 6, 7, and 8 respectively. 
We overplot the predicted radial velocity of the brown dwarf from orbital fits to the relative astrometry from \citet{Crepp2012} and RV measurements of the host star \citep{Crepp2012,Rosenthal2021}. The orbit fits were done with \texttt{orbitize!} \citep{Blunt2020} following its RV tutorial\footnote{\url{https://orbitize.readthedocs.io/en/latest/tutorials/RV_MCMC_Tutorial.html}} and using the \texttt{emcee} \citep{emcee2013} sampler to obtain a posterior of allowed orbits.  This orbital RV of the companion in each epoch is predicted from this orbit fit and is subsequently subtracted from the estimated RV of the planet when running the exomoon search.
Similarly to fitting the centroid of a Gaussian \citep{King1983}, the RV precision goes as the typical linewidth in the spectrum divided by the total S/N of the detection. In the case of HR 7672 B, the large spin with $v\sin i=45.0\pm0.5\,\rm{km.s}^{-1}$ \citep{WangJi2022} is a limiting factor in deriving more precise RVs. The impact on the exomoon sensitivity of other fundamental parameters such as the brightness, age, mass, and separation from the star are discussed in \autoref{sec:generalcase} in the context of TMT/MODHIS.

\subsection{Exomoon sensitivity}
\label{sec:exomoon sensitivity}

The open-source Python package \texttt{RVSearch}\footnote{\url{https://github.com/California-Planet-Search/rvsearch}} \citep{Rosenthal2021} is used to look for possible exomoons around HR 7672 B and derive the sensitivity of our KPIC RV time series.
\texttt{RVSearch}  is a planet search algorithm that was developed by the California Legacy Survey for high-precision radial velocity surveys \citep{Howard2016,Rosenthal2021,Fulton2021}. 
Planets are detected from periodograms, which are expressed as the difference in Bayesian Information Criterion (BIC) between a model including the planet and a model without it \citep{Rosenthal2021}. The $\Delta\mathrm{BIC}$ can be used to select the model that best represents the data, or, in other words, determine if a planet is necessary to explain the observations. Planet candidates are detected by iteratively adding additional planet signal to the model \citep{Rosenthal2021}. For each iterative search, the algorithm fits a detection threshold to the periodogram using the power law noise model described in \cite{Howard2016}. To characterize the search completeness of a dataset, \textit{RVSearch} performs injection-recovery tests, drawing many synthetic planet signals, injecting them in the data, and checking whether their signals surpass the last detection threshold. The simulated signals were injected as described in \citep{Rosenthal2021} with period and $M \sin{i}$ from log-uniform distributions, and eccentricity from an empirically calibrated beta distribution \citep{Kipping2013}.

\begin{figure*}
\gridline{\fig{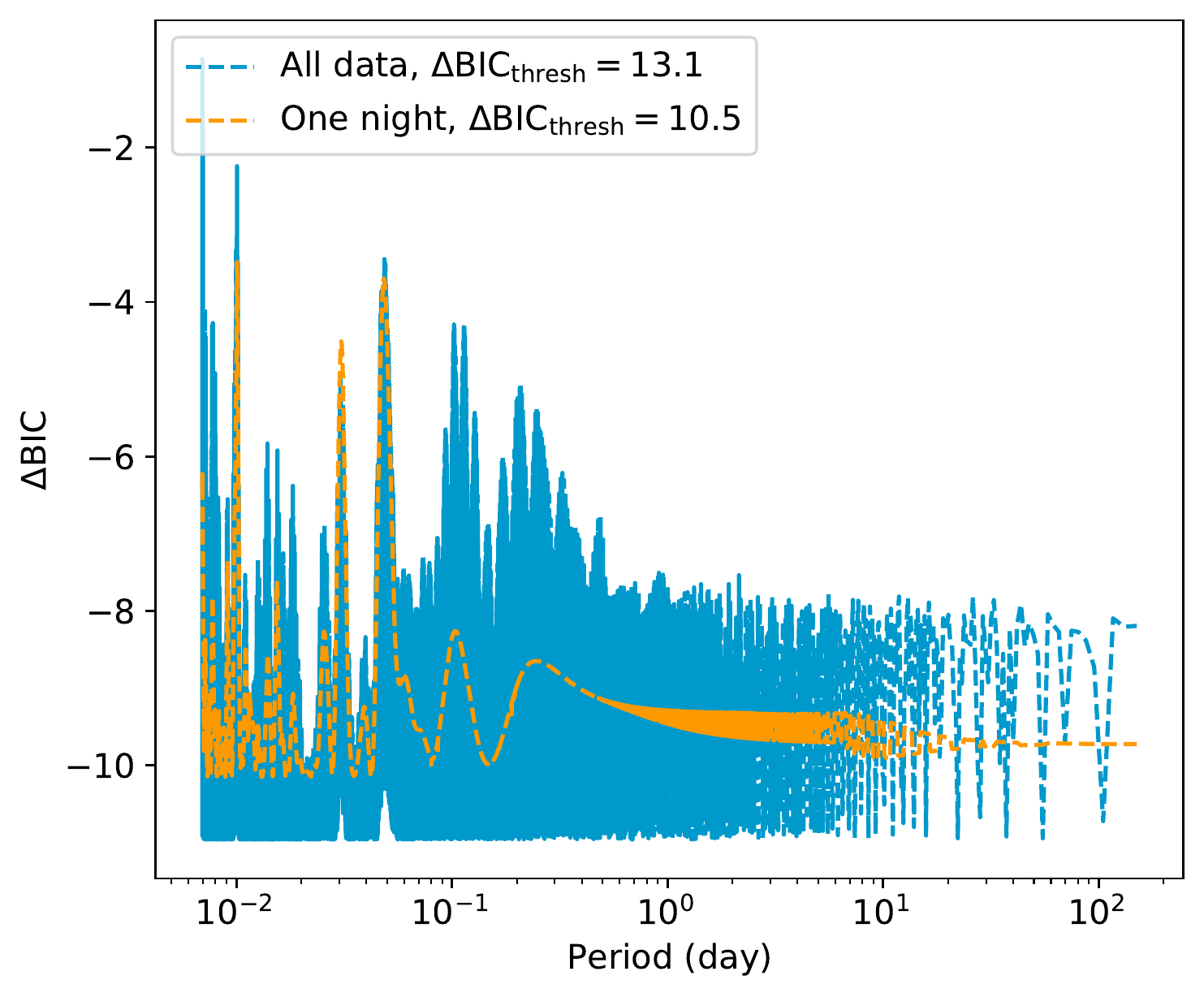}{0.49\textwidth}{(a) Periodogram}
		\fig{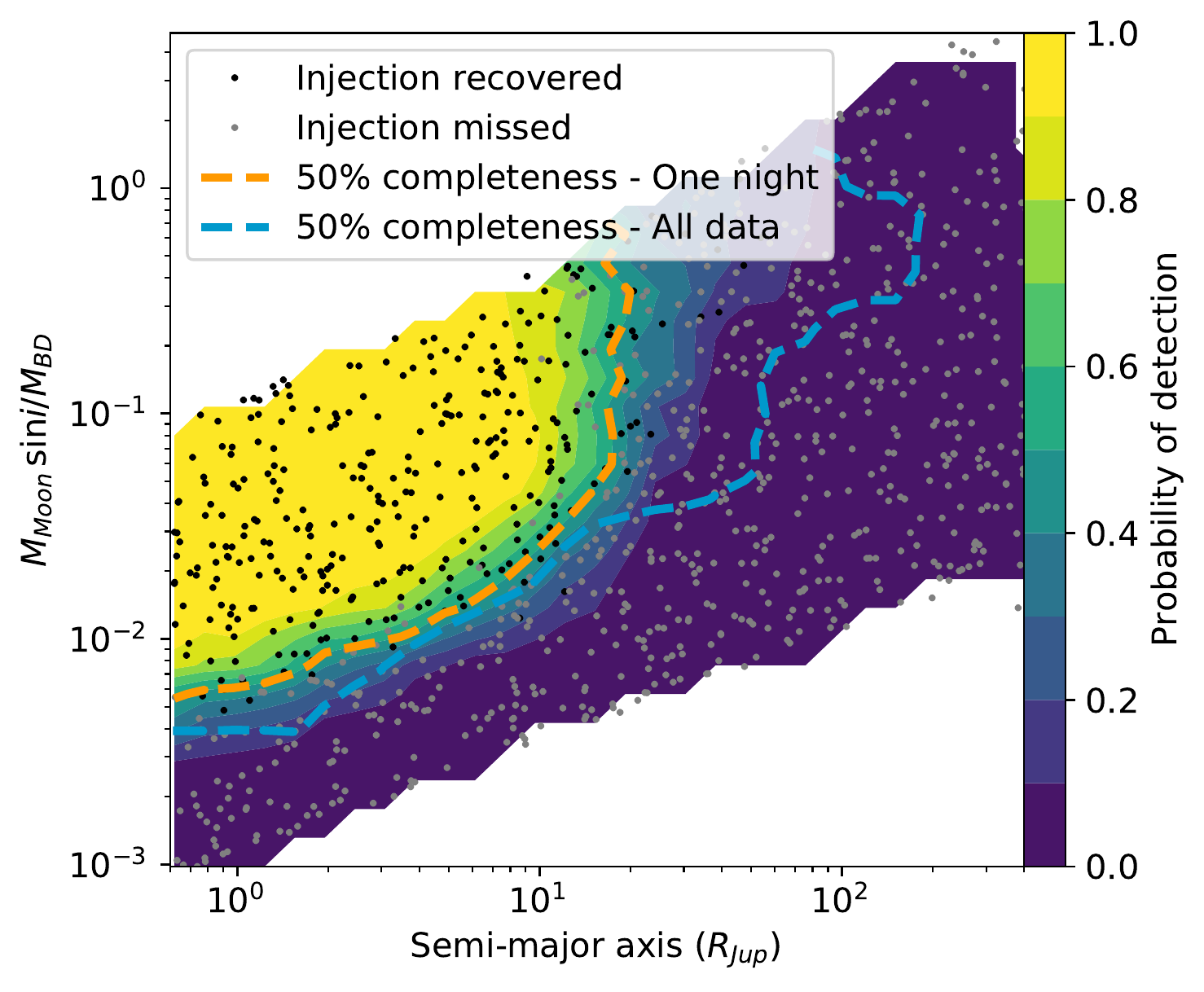}{0.49\textwidth}{(b) Completeness}
          }
\caption{Exomoon detection limits around HR 7672 B with the Keck Planet Imager and Characterizer (KPIC) using the open-source python module \texttt{RVsearch} \citep{Rosenthal2021}. (Left) Periodogram of the RV times series shown in \autoref{fig:rvs} expressed a $\Delta$BIC comparing a model with and a model without a planet. The empirical detection threshold is indicated in the legend. (Right) Exomoon completeness derived from injection and recovery tests. The periodogram and the completeness are shown for two cases: the single full night of observations on 07/04/2022 and all the available data including three additional epochs with 1-2 hours of data each. The variable conditions on 07/04/2022 led to HR 7672 B to not be detected during portions of the night, or in the RV precision to get significantly worse. By simulating RV time series, we estimate that the lost data only affected the final sensitivity by $~20\%$.
}
\label{fig:rvsearch_HR7672B}
\end{figure*}

\texttt{RVSearch} is directly applicable to the search for exomoons by replacing the properties of the star by the ones of the planet. We assume that each spectral order in NIRSPEC has a different zero RV point due to possible inconsistencies between them. This can be done with \texttt{RVSearch}, which linearly solves for offsets between subsets of RVs, and uses a wide, Gaussian, uninformative prior on white noise for each subset. This feature is usually used to fit data from different instruments.
Two analyses are performed, first only using the long night of observations (07/04/2021) and then all the available data. The latter provides a longer time baseline. The resulting periodograms and exomoon completeness are shown in \autoref{fig:rvsearch_HR7672B}. By combining the four epochs, the observations are sensitive to satellites with a mass ratio of $1\%$ at semi-major axes similar to that of Io ($6R_{\rm{Jup}}$) around Jupiter or $4\%$ at the distance of Callisto ($15R_{\rm{Jup}}$). 
While these are encouraging results, the smallest detectable satellites would be as large as Jupiter due to the already large mass of HR 7672 B. As shown in \autoref{sec:generalcase}, targeting smaller brown dwarfs and planets does not generally allow the detection of moons with smaller absolute masses, because the S/N drops faster than the mass of the object due to the decreasing brightness. If satellites around HR 7672 B were to orbit within $\sim10R_{\rm{Jup}}$ of the brown-dwarf, they would likely fall within the Roche radius (See \autoref{fig:moonprospects_RVonly}). Such satellites would be tidally disrupted and likely result in the formation of rings around the planet. It is possible that this issue would prevent the formation of a resonant chain of satellites if the inner edge of the decretion disk falls within the Roche limit. This is for example cited as a possibility to explain the difference between the Galilean and the Saturnian satellite systems in \citet{Batygin2020}.
At the other end of possible satellite semi-major axes, stable orbits can generally exist up to one half of the Hill sphere for prograde orbits \citep{Shen2008}. The Hill sphere of HR 7672 B being $r_H\approx5.6\,\mathrm{au}=1.2\times10^4 R_{\rm{Jup}}$, time series like these ones will not be sensitive to the vast majority of possible orbits without observations spanning years or decades.

\section{Future prospects for HR 7672 B and HR 8799 c}
\label{sec:futurefacilities} 

\subsection{Simulations}

In this section, we simulate observations from current and future instrumentation at the Keck observatory and TMT to estimate the properties of putative satellites that should be detectable using planetary RVs. We use an instrument and observation simulator called \texttt{PSIsim} \footnote{\url{https://github.com/planetarysystemsimager/psisim}}, which was first developed for the Planet Systems Imager \citep[PSI][]{Fitzgerald2022} instrument concept for TMT, and then expanded to include other instruments and telescopes. \texttt{PSIsim} is first used to estimate the RV precision. Then, RV times series are simulated assuming 6 full nights of observations over 25 days, and the exomoon sensitivity is finally computed using \texttt{RVSearch}. These simulations are meant to represent an ideal scenario in terms of instrument performance and telescope time allocation.

We simulate observations of two substellar companions, the brown-dwarf companion HR 7672 B and the planet HR 8799 c, with four generations of instruments. 
An exhaustive analysis of all directly imaged companions is beyond the scope of this work so HR 8799 c was chosen as a representative example of the field with a planetary mass. HR 8799 is also the only other high-contrast system with published RV time series and exomoon upper limits \citep{Vanderburg2021}.
The four instruments considered in this work are Keck/KPIC I, Keck/KPIC II, Keck/HISPEC, and TMT/MODHIS. KPIC I corresponds to observations carried out pre-2022A \citep{Delorme2021b}. KPIC II refers to the series of upgrades started during the first semester of 2022 with the primary goal of doubling the instrument throughput \citet{Jovanovic2020,Echeverri2022}. The High-resolution Infrared Spectrograph for Exoplanet Characterization (HISPEC) is expected to provide Y-K ($0.98-2.46\,\mu\mathrm{m}$) spectroscopy at a spectral resolution of $R>100,000$ \citep{Mawet2019}. The Multi-Object Diffraction-limited High-resolution Infrared Spectrograph (MODHIS) is a similar instrument to HISPEC planned for the future TMT. A broader range of exoplanet masses is explored in \autoref{sec:generalcase} for this latter TMT instrument.

\texttt{PSIsim} includes full budgets of the throughput and thermal background for each instrument, telescope, and the Earth atmosphere. The Strehl ratio is calculated based on a empirically calibrated model of the adaptive optics' performance under median seeing conditions for Maunakea. For KPIC I and KPIC II, we assumed Keck AO's current performance with the infrared Pyramid Wavefront Sensor described in \citet{Bond2020}. For HISPEC, we assumed extreme-AO performance as predicted for the upcoming HAKA high-density deformable mirror upgrade (W.M. Keck Observatory, private communication). The star is modeled with a $PHOENIX$ model \citep{Husser2013} and the substellar companion with a BT-Settl atmospheric model grid\footnote{\url{https://phoenix.ens-lyon.fr/Grids/BT-Settl/CIFIST2011c/}} \citep{Allard2012}. \autoref{tab:psisimparas} includes the input parameters and the predicted RV precision for these simulations. The simulations include a level of systematics at 1\% of the continuum, which is modeled by an additional white Gaussian noise. Otherwise, the estimated RV precision assumes a perfect data reduction.  

The predictions from PSI-sim are about a factor two more sensitive than existing measurements with KPIC I (See \autoref{tab:psisimparas}). This difference can first be explained by uncorrected wavefront errors reducing the throughput, both non-common path aberrations and uncorrected atmospheric turbulence. Then, our current data analysis framework remains limited in its ability to model KPIC systematics. As explained in \autoref{sec:FM}, only the redder orders of NIRSPEC are being reduced due to strong telluric lines in the bluer orders, and an imperfect Fourier filtering is used to remove the fringing. The gap between the simulations and the measurements should decrease as observing strategies and data reduction frameworks are improved.

The final expected exomoon sensitivity of the four instruments is shown for the two companions in \autoref{fig:moonprospects_RVonly}. 
For a fixed time sampling of the RV series, the minimum detectable mass ratio is approximately proportional to the RV semi amplitude of the signal, which is also proportional to the RV precision of the instrument, so the improvement for each generation of instrument can be read from the simulated RV precision shown at the bottom of \autoref{tab:psisimparas}.
These simulations are compared to other detection techniques in \autoref{app:comparison}, specifically astrometric monitoring of the companion or spatially resolving the moon through imaging. We separately discuss the possibility of detecting transiting exomoons using the Rossiter-McLaughlin (RM) effect in \autoref{sec:RM}.

\subsection{Comparing to solar system moons}

\begin{figure*}
\gridline{\fig{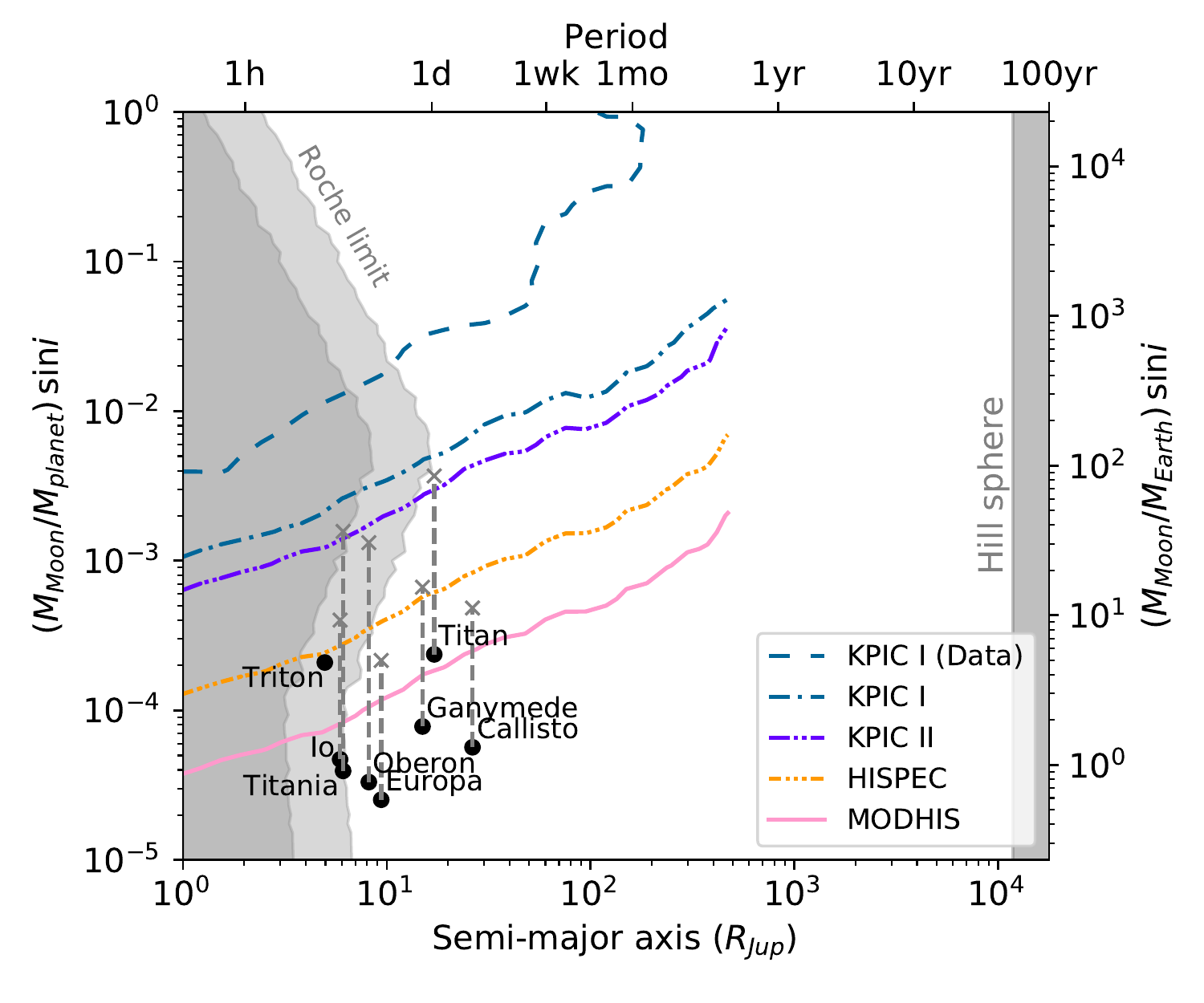}{0.49\textwidth}{(a) HR 7672 B}
		\fig{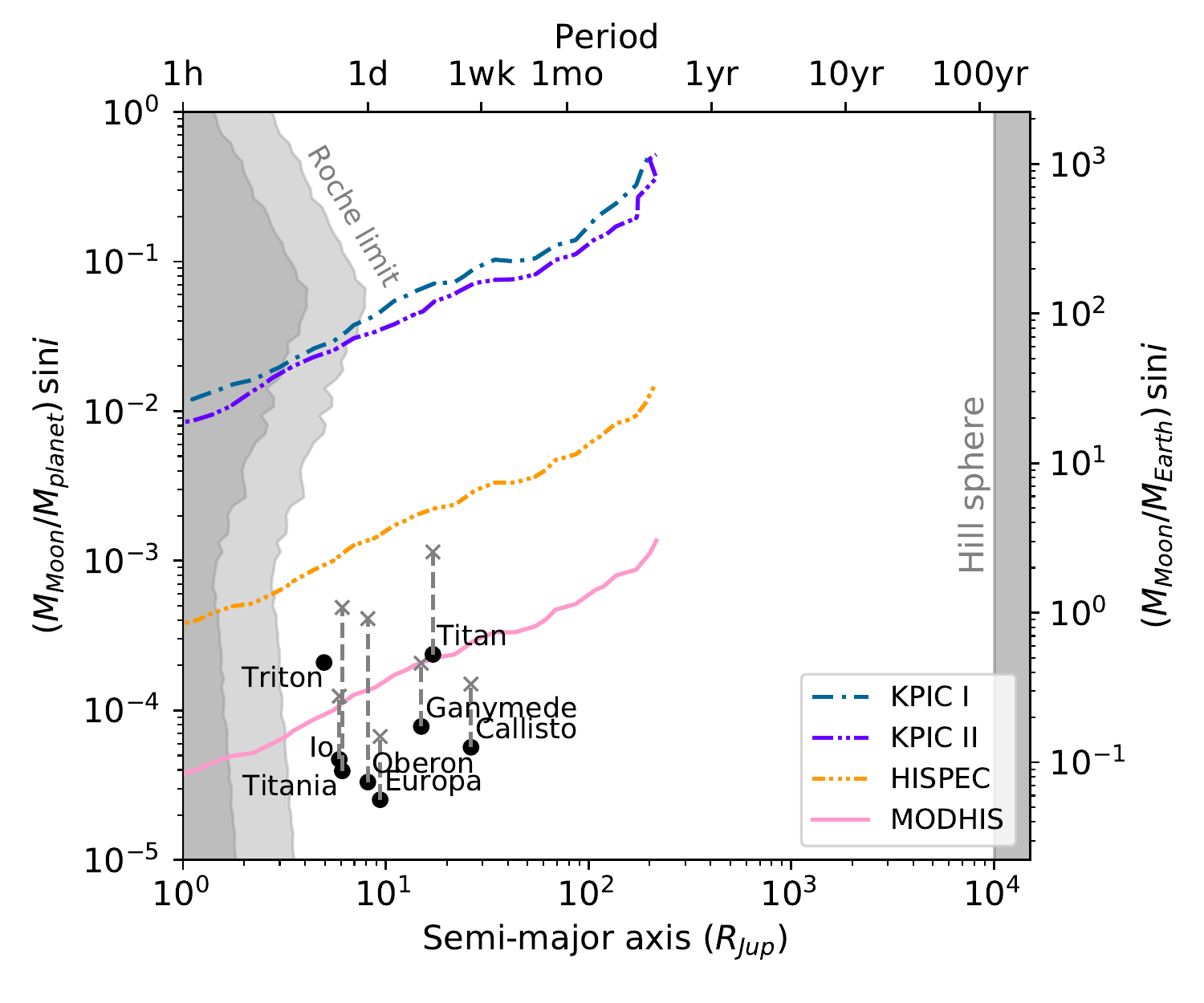}{0.49\textwidth}{(b) HR 8799 c}
          }
\caption{
Future prospects for exomoon detections around the brown dwarf companion HR 7672 B (left) and planet HR 8799 c (right). Simulated sensitivity for Keck/KPIC I, Keck/KPIC II, Keck/HISPEC, and TMT/MODHIS are shown in colored curves assuming 6 nights of observations over 25 days. The sensitivity demonstrated in this work from $\sim1.5$ nights of KPIC observations is labeled as KPIC I (Data).
The mass ratios of the Galilean satellites are shown as black dots for comparison. Their predicted scaled-up mass ratios, $q$, accounting for the larger mass, $M$, of the brown dwarf compared to Jupiter are shown as grey crosses ($q\propto \sqrt{M}$; \citet{Batygin2020}). The Roche limit is computed for both a rigid and a fluid satellite shown as the inner and outer greyed region respectively.
}
\label{fig:moonprospects_RVonly}
\end{figure*}

The mass ratios of the largest gas giant satellites in the solar system are also shown in \autoref{fig:moonprospects_RVonly} for comparison. The higher planet masses, $M$, of directly imaged planets and brown dwarfs compared to the solar system could yield significantly bigger moons, so we also include scaled-up mass ratios, $q$, according to $q\propto \sqrt{M}$ \citep{Batygin2020}.
While the CPD does scale with the Hill Sphere, we do not expect the semi-major axis of satellites to depend on this parameter. Indeed, young moons are thought to migrate toward the planet during their formation due to the interaction with the gas. The migration is stopped at the inner radius of the CPD which is set by the magnetic field of the planet \citep{Batygin2020}. In this work, we therefore keep the semi-major axis of the solar system satellites constant. A caveat is that large moons could be susceptible to tidal forces if they form or migrate too close to the planet within the Roche limit. The Roche limit is calculated using the mass-radius relationship from \citet{Chen2017} and their associated Python package\footnote{\url{https://github.com/chenjj2/forecaster}}. However, this relationship does not account for the fact that young objects are likely inflated.

\begin{table}[ht]
  \centering
\begin{tabular}{ lcc } 
 \hline
\multicolumn{3}{c}{Parameters} \\
 \hline
Star - Phoenix model  & HR 7672 & HR 8799\\
 \hline
Apparent K mag & 4.4\tablenotemark{a} & 5.2\tablenotemark{a} \\
Effective temperature ($T_{\mathrm{eff}}$) & $6000\,\mathrm{K}$\tablenotemark{b} & $7400\,\mathrm{K}$\tablenotemark{c} \\
Surface gravity ($\mathrm{log}(g)$) & 4.5\tablenotemark{b} & 4.5\tablenotemark{c} \\
Spin ($v\mathrm{sin}(i)$; km/s) & 5.6\tablenotemark{d} & 49\tablenotemark{e} \\
 \hline
 Companion - BTsettl model & HR 7672 B & HR 8799 c\\
\hline
Mass & $73M_{\mathrm{Jup}}$\tablenotemark{b} & $7M_{\mathrm{Jup}}$\tablenotemark{f} \\
Apparent K mag & 13.0\tablenotemark{b} & 16.1\tablenotemark{g} \\
Effective temperature ($T_{\mathrm{eff}}$) & $1800\,\mathrm{K}$\tablenotemark{b} & $1200\,\mathrm{K}$\tablenotemark{h} \\
Surface gravity ($\mathrm{log}(g)$) & 5.5\tablenotemark{b} & 4.0\tablenotemark{h} \\
Spin ($v\mathrm{sin}(i)$) & $45\,\mathrm{km.s}^{-1}$\tablenotemark{b} & $10\,\mathrm{km.s}^{-1}$\tablenotemark{i}\\
Separation & $0.72^{\prime\prime}$\tablenotemark{j} & $0.95^{\prime\prime}$\tablenotemark{j} \\
 \hline
\multicolumn{3}{c}{Telescope and instrument} \\
 \hline
airmass & 1.2&\\
water vapor column & $1.5\,\mathrm{mm}$&\\
integration time ($t_{\textrm{int}}$) & $5\,\textrm{min}$ &\\
 \hline
 \hline
\multicolumn{3}{c}{Predicted RV sensitivity ($\textrm{m.s}^{-1}$)} \\
\multicolumn{3}{c}{assuming $0.6^{\prime\prime}-1.0^{\prime\prime}$ seeing} \\
 \hline
 & HR 7672 B & HR 8799 c\\
 \hline
  Keck/KPIC I (measured) & $\sim2,000$\tablenotemark{k} & $\sim7,000$\tablenotemark{i} \\
  Keck/KPIC I (simulated) & 800-1400 & 3,000-5,000 \\
  Keck/KPIC II & 500-800 & 2,000-3,000\\
  Keck/HISPEC & $\sim200$ & 100-200\\
  TMT/MODHIS & 30-40 & 10-20 \\
 \hline
\end{tabular}
\caption{Radial velocity (RV) precision simulations of current and future instrumentation for two substellar companions: HR 7672 B and HR 8799 c. (Top) Representative parameters for the telescope, instrument, star, and companions used in the \texttt{PSIsim} simulations. (Bottom) Predicted RV sensitivity for values of seeing ranging from $0.6^{\prime\prime}$ to $1.0^{\prime\prime}$.
}
\tablerefs{a: \citet{Cutri2003}, b: \citet{WangJi2022}, c: \citet{WangJi2020}, d: \citet{Luck2017}, e: \citet{Royer2007}, f: \citet{Wang2018}, g: \citet{Currie2011}, h: \citet{Wang2018}, i: \citet{Wang2021}, j: \url{http://whereistheplanet.com/} \citep{whereistheplanet}, k: This work}
\label{tab:psisimparas}
\end{table}

\section{Future exomoon sensitivity of TMT/MODHIS}
\label{sec:generalcase}

Looking to the future, we expect substantial gains in RV precision by using the next generation of high-resolution spectrographs on large telescopes. These gains in RV precision will lead to enhanced sensitivity to systems with lower mass, close in exomoons, which would form in a similar way to the Galilean moons around Jupiter. 

\begin{figure*}
\gridline{\fig{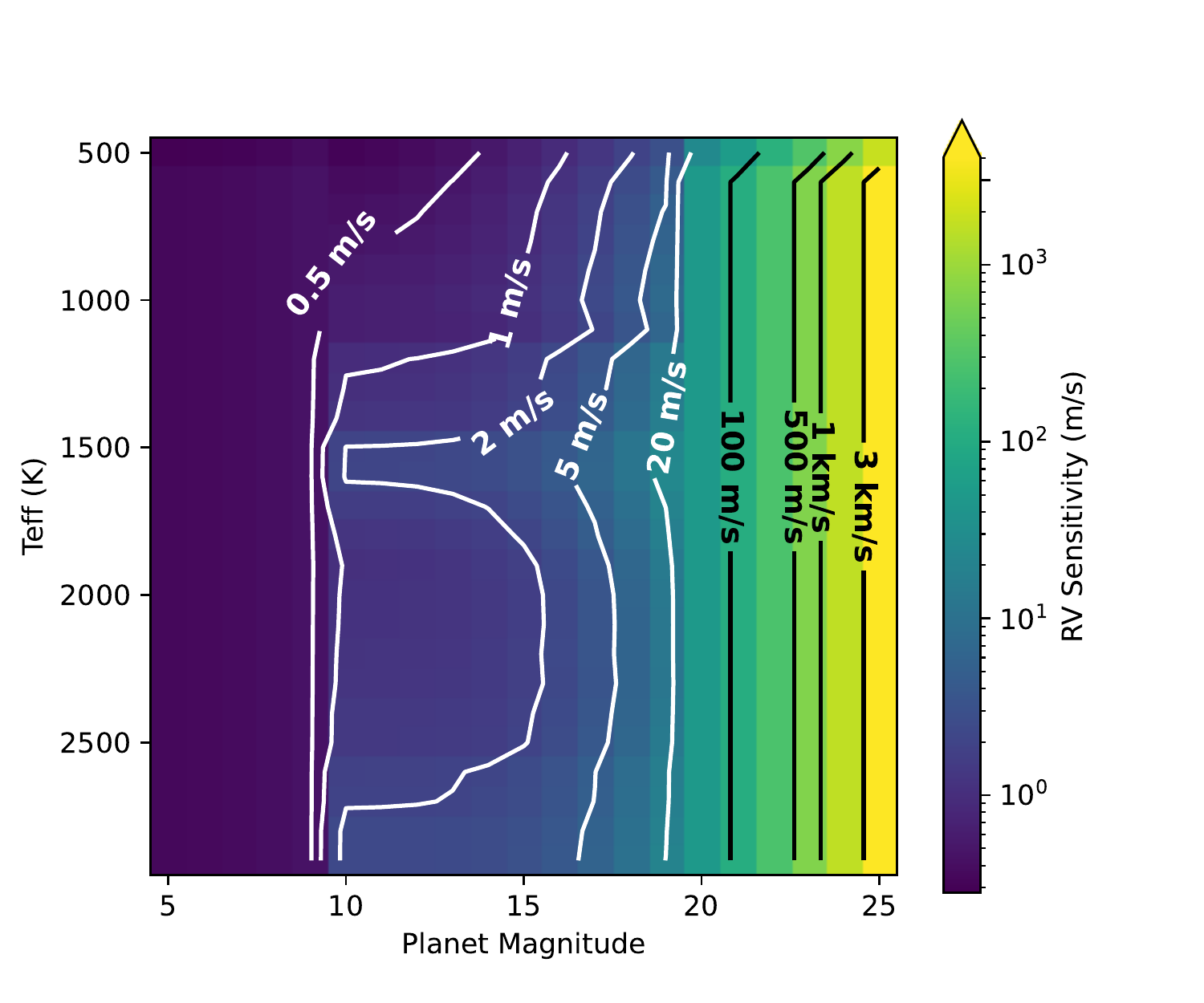}{0.49\textwidth}{(a) RV sensitivity}
		\fig{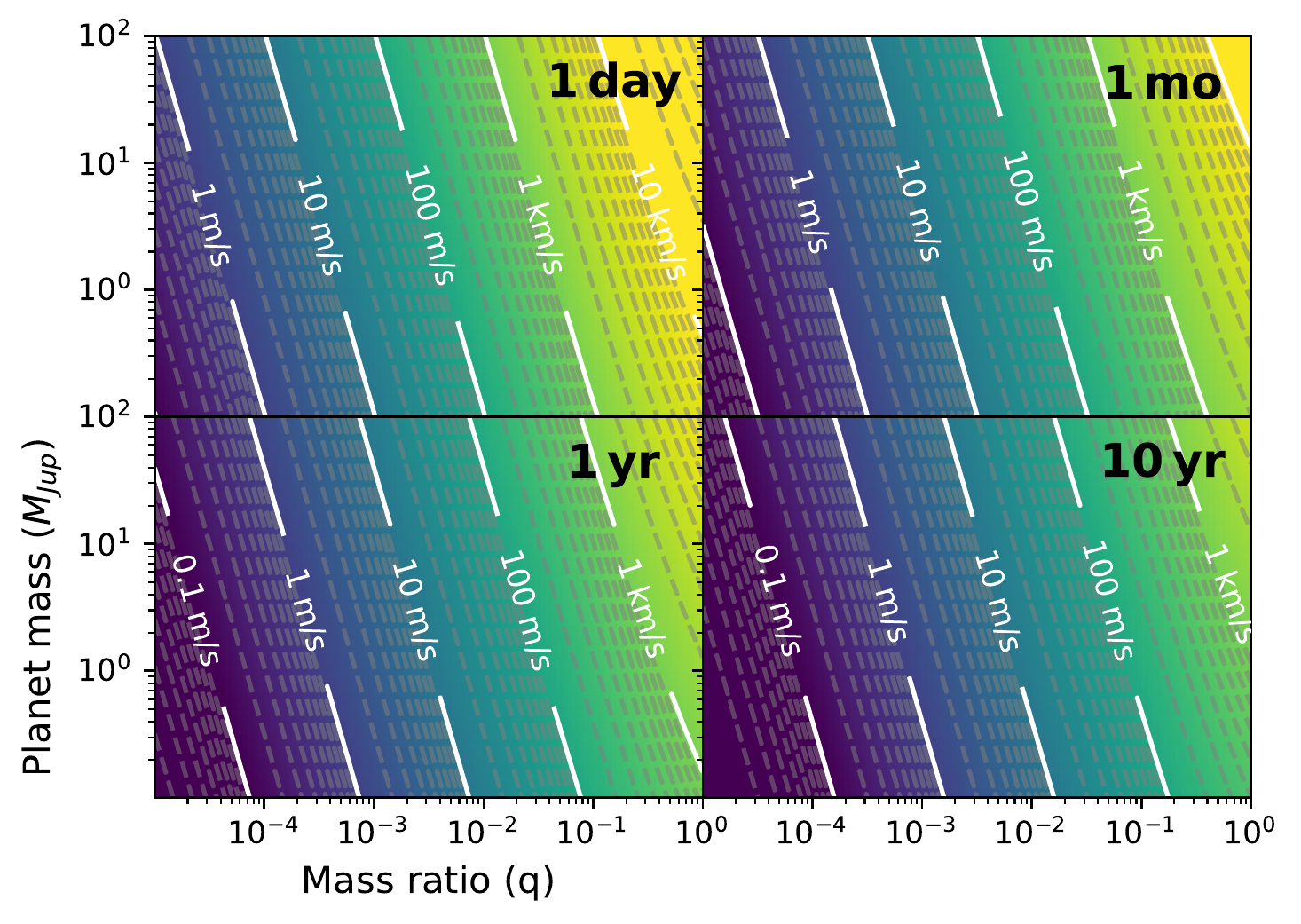}{0.52\textwidth}{(b) RV semi-amplitude}
          }
\caption{RV precision of MODHIS. (Left) The RV sensitivity of MODHIS for model planets around a HR 8799-like host star using BT-Settl models \citep{Allard2012}. The RV sensitivity was predicted using \texttt{PSIsim} for a single, two hour exposure. 
Both the contour curves and color map indicate the RV sensitivity for a specified effective temperature and apparent magnitude of the model planet. The RV sensitivity relies more on the brightness of the planet than its effective temperature. However, the RV sensitivity decreases for planets with temperatures between 1500 and 1700 K due to the L-T transition. (Right) The RV semi-amplitude for different planet masses and mass ratios. Note, increasing the exposure time will increase the RV sensitivity.}
\label{fig:mag_teff}
\end{figure*}

Using the same framework as in \autoref{sec:futurefacilities}, we calculate the RV sensitivity for a variety of simulated planets that could exist around a host star with the properties of HR 8799 referenced in \autoref{tab:psisimparas}. We modeled planets with varying effective temperatures and apparent magnitudes, fixing the separation between the planet and star to 700 mas and the surface gravity of the planet to $log(g)=4.5 \mathrm{cm.s}^{-2}$, and used \texttt{PSIsim} to calculate the RV sensitivity.
The effect of the starlight contamination on RV sensitivity can be neglected for the type of directly imaged planets that are known today and would be observed with TMT. The RV sensitivity vary by less than 20 percent for planets that lie beyond 500 mas and have a flux ratio greater than $\sim3 \times 10^{-6}$. On average, for every 0.5 dex change in surface gravity on the planet, the RV sensitivity changes by $\pm 0.7$ m/s. \autoref{fig:mag_teff} (a) shows the RV sensitivity MODHIS could have for a single, two hour exposure, for planets of varying effective temperatures and apparent magnitudes around an HR 8799 like star. The RV sensitivity of MODHIS driven by the brightness of the planet more than than its temperature. However, the RV sensitivity is decreased for planets with temperatures between 1500 and 1700 K using the BT-settl model grid due to the L-T transition. At these temperatures, clouds form in the upper layers of the atmosphere, shrouding detectable spectral lines. For a given planet temperature and magnitude, the RV precision of TMT/MODHIS \autoref{fig:mag_teff} (a) can be compared to the RV semi amplitude  in \autoref{fig:mag_teff} (b) as a function of the planet mass, the mass ratio, and the period of the satellite. However, such a comparison assumes multiple epochs of observations with a given sensitivity in order to detect a moon with a similar RV semi amplitude.  

\begin{figure*}
\gridline{\fig{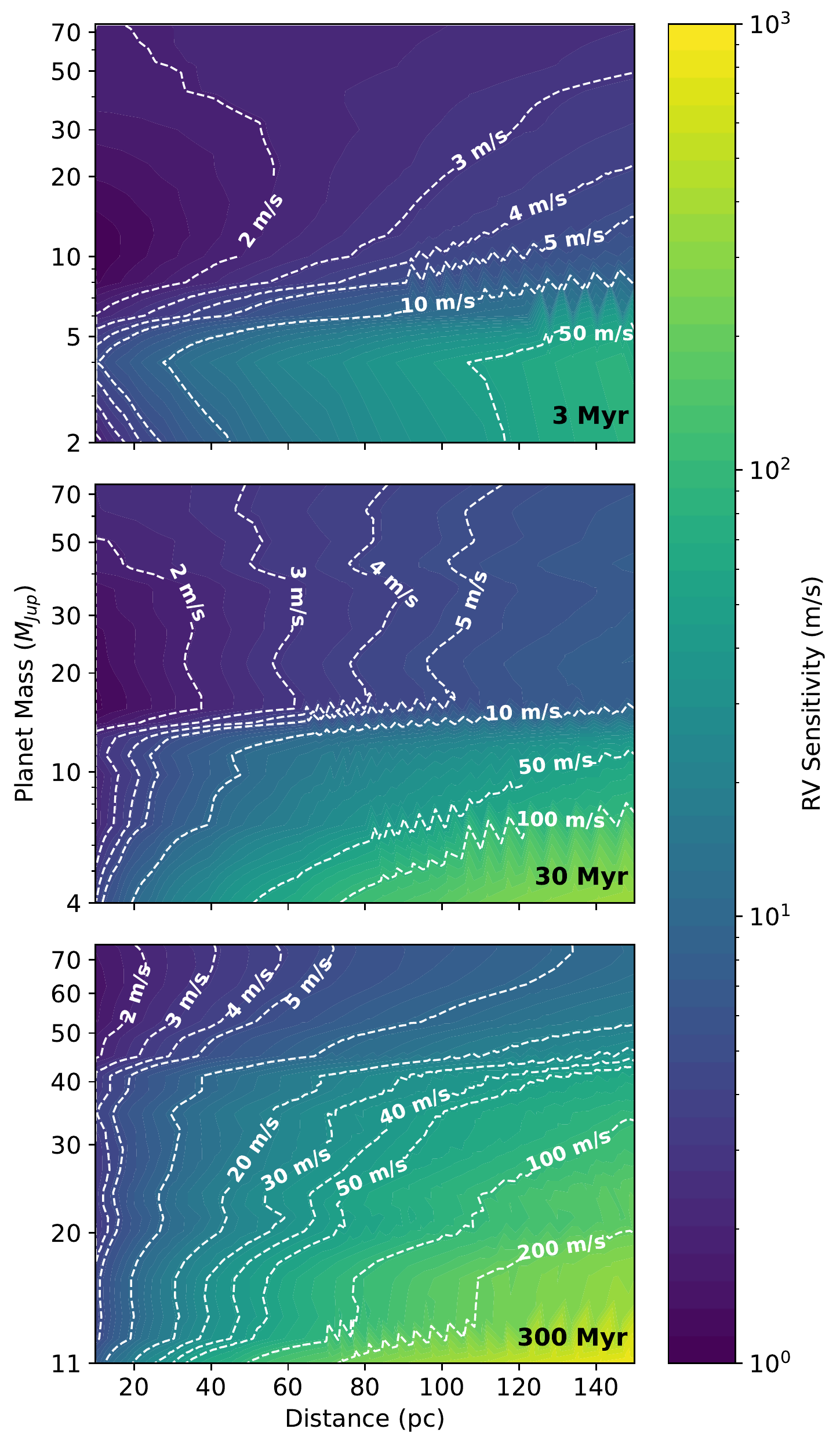}{0.48\textwidth}{(a) RV sensitivity}
		\fig{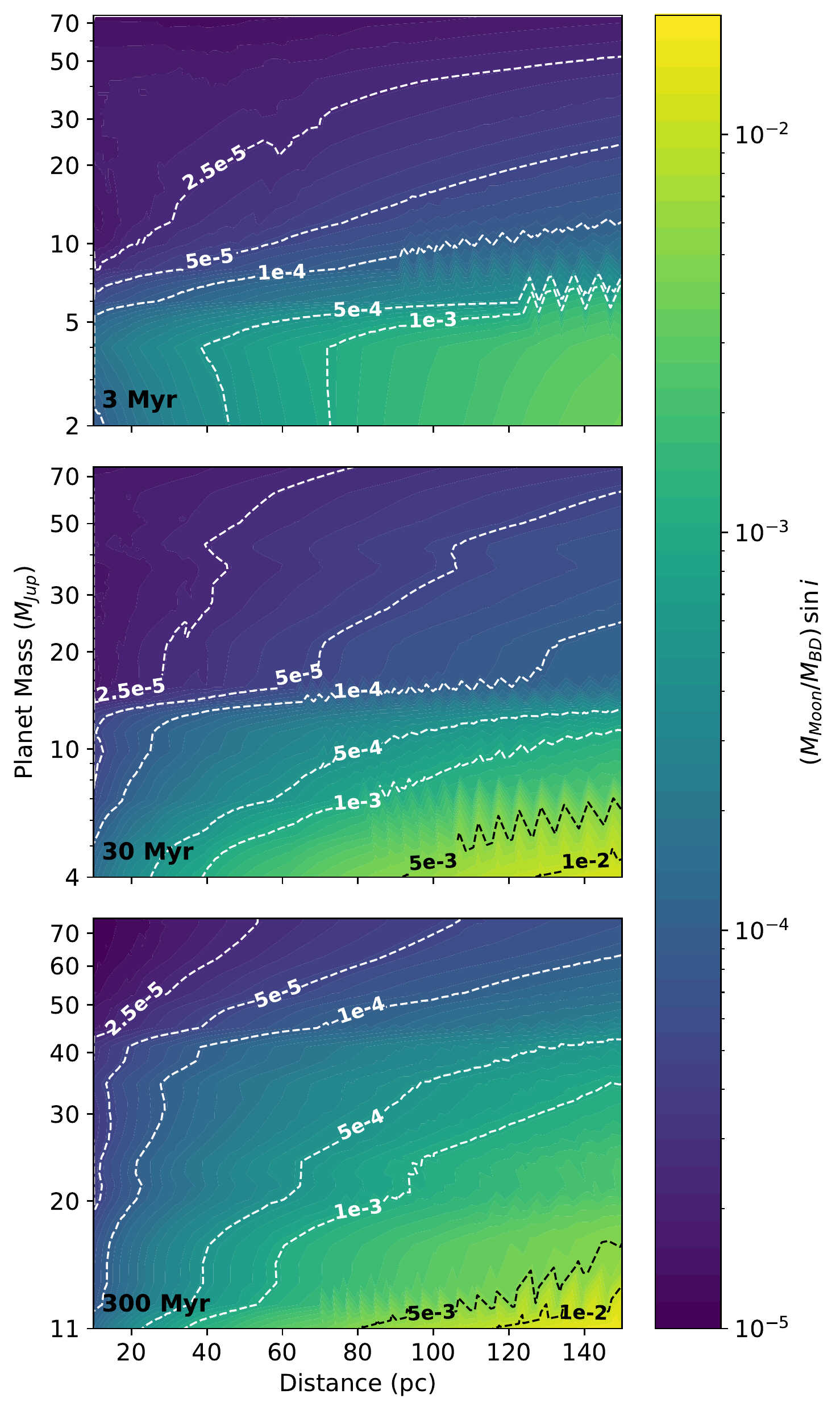}{0.49\textwidth}{(b) Detectable mass ratio}
          }
\caption{RV precision and detectable mass ratio of MODHIS similar to \autoref{fig:mag_teff}, but as a function of planet mass,  distance, and age of the system. (Left) BT-Settl evolutionary models \citep{Allard2012b} were used to infer the mass of the planet and distance to the system at an age of 3, 30, and 300 Myr. The minor contour lines cover an evenly spaced, 50 step log scale from 0 to 1 km/s. RV sensitivity decreases the further the system is away and the lower in mass the planet is. The large decrease in RV sensitivity when the companion mass is below $\sim13$ $M_{Jup}$ for young systems is due to the difference in cooling rates between brown dwarfs and planets over time. (Right) The mass ratio detectable by MODHIS assuming a fixed semi major axis for the moon equal to that of Callisto ($\approx26R_\mathrm{Jup}$). For each planetary mass and distance from panel (a), we create an RV time series assuming six nights of observations over 25 days with error bars that represent the RV sensitivity calculated by \texttt{PSIsim}.
}
\label{fig:dist_mass}
\end{figure*}

In the following, the surface gravity, temperature, and mass of the planet are treated more self-consistently using BT-Settl evolutionary grids \citep{Allard2012b}. The dependence of the exomoon sensitivity to the number of observations is also made explicit by using simulated RV time series. 
We therefore express the RV precision and exomoon sensitivity as a function of planet mass and distance to the Sun in \autoref{fig:dist_mass}. We fixed the age of the system to different values to represent the parameter space occupied by different populations of stars. The 3 Myr age group is representative of the youngest stars, such as those found in star forming regions (e.g. Ophiuchus, Taurus, etc). The 30 Myr age group is representative of young moving groups, such as Beta Pictoris Moving Group and the Tucana and Horologium Associations. The 300 Myr age group is representative of the oldest directly imaged substellar companions. The RV sensitivity decreases the further the system is away at each distinct age. For younger systems, there is larger decrease in sensitivity as the mass of the planet decreases below $\sim13$ $M_{Jup}$. The large decrease in RV sensitivity once the object is below $\sim13$ $M_{Jup}$ is due to the onset of deuterium burning for brown dwarfs, which makes them much more luminous than a planet of a similar mass. Another interesting feature in \autoref{fig:dist_mass} (a) is the apparent independence of the RV precision to the brown dwarf mass above $\sim13$ $M_{Jup}$ at 30 Myr. This can be explained by the facts that the RV precision is mostly driven by the brightness of the object, and that brown dwarfs have a similar brightness over a range of masses around this age. Indeed, larger brown dwarfs cool faster than smaller ones resulting in the different cooling curves to meet over a small range of brightness around 30 Myr as illustrated in Figure 7 in \citet{Burrows1997}.

\autoref{fig:dist_mass} (b) shows the moons that could be detected around a planet from \autoref{fig:dist_mass} (a) if they were placed at the distance of Callisto. For each planetary mass and distance, we create an RV time series assuming six full eight-hour nights of observations over 25 days with error bars that represent the RV sensitivity calculated by \texttt{PSIsim}. The detection threshold was computed from simulated data created by \texttt{RVsearch} as in \autoref{sec:futurefacilities}. For more massive planets and brown dwarfs, we expect TMT/MODHIS to reach the RV sensitivity needed to look for close in moons with mass ratios smaller than $10^{-4}$ around brown-dwarfs, similar to the ones found in the solar system for a median age of 30 Myr. However, to detect moons around lower mass, directly imaged planets of the same age, we are sensitive to mass ratios of $10^{-3}$ or larger. 

\section{Discussion}
\label{sec:discussion}

\subsection{The viability of exomoon RV searches}

Using KPIC, we derive the most sensitive upper limits on the mass ratio of satellites orbiting a high-contrast substellar companion. We rule out satellites larger than 1-4\% the mass of the brown dwarf HR 7672 B at separations similar to the Galilean moons. 
Based on end-to-end simulations, we predict that instruments such as TMT/MODHIS could be two orders of magnitude more sensitive. This would be sufficient to detect moons forming in the CPD of a planet with mass ratios of $\sim10^{-4}$, albeit with a substantial investment in observing time. If the satellite to planet mass ratio grows as $q\propto\sqrt{M}$, with $M$ the mass of the planet, the Keck/HISPEC should be sensitive to these objects around brown dwarfs. Any detection with HISPEC, or lack thereof, will therefore already be capable of constraining CPD formation models. 
In order to validate our instrument simulations, we compared them with existing observations. The gap in sensitivity can be explained by imperfections in the data reductions. A continued investment in more accurate data processing algorithms or observing strategies is therefore required in order to realize these predictions. Planet variability will also be a challenge to overcome using the different timescales and the wavelength dependence of the variability compared an exomoon signal for example \citep{Vanderburg2018}. Measuring the variability of substellar companions would in fact be an important result of exomoon surveys to better understand the physics of their atmospheres \citep{Biller2017}.

Binary formation processes favor high-mass ratios so they would be more easily detectable than the smaller satellites forming by accretion in the CPD. The majority of multiplicity surveys for isolated brown dwarfs \citep{Fontanive2018}, or companion brown dwarfs  \citep{Burgasser2005,Lazzoni2020}, have searched for visual companions, leaving the separation regime of $<1\,\mathrm{au}$ underexplored. \autoref{fig:mag_teff} (b) shows that unresolved binary substellar companion would be detectable with RV precision between $0.1-1\,\mathrm{km.s^{-1}}$, which is already routinely achieved with KPIC. As an example, the measured dynamical mass of the brown dwarf companion HD 47127 B suggest that it could be a binary \citep{Bowler2021}, but this specific companion is too faint ($K\sim18.4$) to be a practical target for KPIC.

From \autoref{app:comparison} and \autoref{fig:moonprospects_comparison}, we conclude that the different detection techniques are sensitive to distinct regions of the parameter space, and therefore complementary, not unlike exoplanet searches. If exomoons follow the model of solar system gas giant satellites, RV searches could be the most promising approach due to its sensitivity to short period moons. However, unless the theoretical prediction that bigger planets form even bigger moons hold true, small satellites with mass ratios $\sim 10^{-4}$ might only be detectable around brown dwarfs.

\subsection{Detections of moons using the Rossiter-McLaughlin effect}
\label{sec:RM}

As suggested in \citet{heller2014}, an alternative strategy to look for exomoons around directly imaged planets using RV measurements could be to look for transiting moons through the Rossiter-McLaughlin \citep[RM; ][]{Gaudi2007} effect on the planet.
Precise photometric calibration and stability of high-constrast instrument is notoriously difficult \citep{Wang2022}, so detecting a RM event during a transit could be easier than detecting its photometric counterpart.

An RM event consists of the subsequent masking of a portion of the blue and red-shifted areas of the surface of a spinning object, therefore leading to large and very distinct deviations of the measured RV. The amplitude of the RV signal can be hundreds of times larger than the RV semi amplitude due to the orbital motion of the moon. Its amplitude is proportional to the spin of the planet, which could make it an interesting alternative to detect the smallest moons around rapidly rotating planets and brown dwarfs. Indeed, the RV uncertainties scale with the spin of the object so detecting the orbital signal of small exomoons could be more challenging. 

The Galilean moons have rather small orbital periods from days to weeks. 
Assuming a random inclination distribution, the transit probability of a moon ($P$) is given by the ratio of the planet radius ($R_p$) and the moon semi-major axis ($d_m$), $P=R_p/d_m$ \citep{Borucki1984}.
Therefore, the probability of a transit of a moon at the separation of Io around Jupiter is 1:6, and 1:27 for the farthest Galilean moon Callisto. Assuming a full 8 hour night of observations, we estimate the probability of observing an RM event for Galilean-like moons around a Jupiter like planet to be around 3\% for Io, 1\% for Europa, 0.3\% for Ganymede, and 0.07\% for Callisto.
However, the orbital periods of the moons would be even shorter around larger substellar companions, which would increase the probabilities up to 17\% for Io, 8\% for Europa, 2.6\% for Ganymede, and 0.6\% for Callisto .
The transits would last between $\sim$2-5 hours for the Galilean moons around Jupiter, but they would only last 15-30 minutes for similar moons around HR 7672 B.

As an example, a satellite around HR 7672 B with a mass of $1M_{\Earth}$ ($q=5\times10^{-5}$) would generate a RM signal of $\sim300\,\mathrm{m.s^{-1}}$ compared to the $\sim0.5\,\mathrm{m.s^{-1}}$ generated by the orbital motion \citep{Gaudi2007}. The amplitude would be $\sim5\,\mathrm{km.s^{-1}}$ for a Neptune-size moon. Multiple satellite systems would increase the probability of a detection.
Given the low detection probability, RM searches could be carried out in synergy with other science cases such as brown dwarf variability \citep{Biller2017}. For example, Doppler spectroscopy also favors long observations of rapidly rotating objects, which would make for ideal datasets for exomoon RM searches.

\subsection{Searching for Pandora: Habitable exomoons}

Estimating the occurrence rate of Earth-sized exoplanets in the habitable zone (HZ) of Sun-like star, called $\eta_{\Earth}$, has been an important goal of exoplanet surveys. While such planets remain challenging to detect, the best estimates of $\eta_{\Earth}$ range between $5-50\%$ to date \citep{Gaudi2021}. However, these are not the only Earth-sized objects that could harbor life in the HZ of their stars. Any rocky satellites orbiting HZ gas giant planets could also provide suitable conditions for life. Close-in exomoons can be protected from stellar radiation by the strong magnetic field of Jovian mass planets \citep{Heller2013}.

 Integrating the distribution of gas giants with an incident flux between $0.3-1.5$ times the solar irradiance on Earth for an optimistic habitable zone, or $0.3-1$ for a conservative habitable zone \citep{Kasting2013}, yields about $5-7$ giant planets per hundred FGKM stars. This is using the giant planet ($30-6000M_{\Earth}\sin{i}$) occurrence rates derived from the California Legacy Survey as a function of stellar irradiation \citep[figure 11; ][]{Fulton2021}. Given that each planet can have multiple satellites, this could represent a significant number of habitable Earth-size moons that are not accounted for in $\eta_{\Earth}$. The occurence rate of habitable exomoons could be constrained by measuring the population of satellites around more distant directly imaged planets and brown-dwarfs.

\section{Conclusion}
\label{sec:conclusion}

In this work, we aimed at evaluating the prospects for radial velocity (RV) detections of exomoons around self-luminous directly-imaged planets. We used real observations as well as end-to-end simulations of future facilities at the Keck observatory and the Thirty Meter Telescope (TMT). 
Using data from KPIC, we were able to derive upper limits for satellites orbiting the brown dwarf companion HR 7672 B at a mass ratio of $1-4\%$ for separations similar to the Galilean moons. Current instrumentation is already sensitive to unresolved binary companions that could form through gravitational instability. We demonstrate that future thirty-meter class telescopes will likely push the sensitivity down to the mass ratios of solar system satellites ($\sim10^{-4}$), which are thought to form in a circumplanetary disk. We note that second generation instruments like Keck/HISPEC on current ten meter class telescopes might be sufficient to detect these moons if theoretical predictions that larger planets form even larger moons hold true. Everything else being equal and considering the RV signal from the orbital motion of the moon, the deepest exomoon sensitivity will be reached for the brightest substellar companions with the smallest spin.
Small moons could also be detected from their Rossiter-McLaughlin (RM) effect on the planetary RV signal. An RM event can be orders of magnitude larger than the orbital signal, albeit with percents level detection probability assuming a full night of observation.
We conclude that the detection of exomoons from planetary RV surveys is now becoming a reality thanks to the development of high-resolution spectrographs dedicated to directly imaged planets.

%% file: appendix.tex
\clearpage
\appendix

\section{Comparing to other detection methods}
\label{app:comparison}

\begin{figure*}
\gridline{\fig{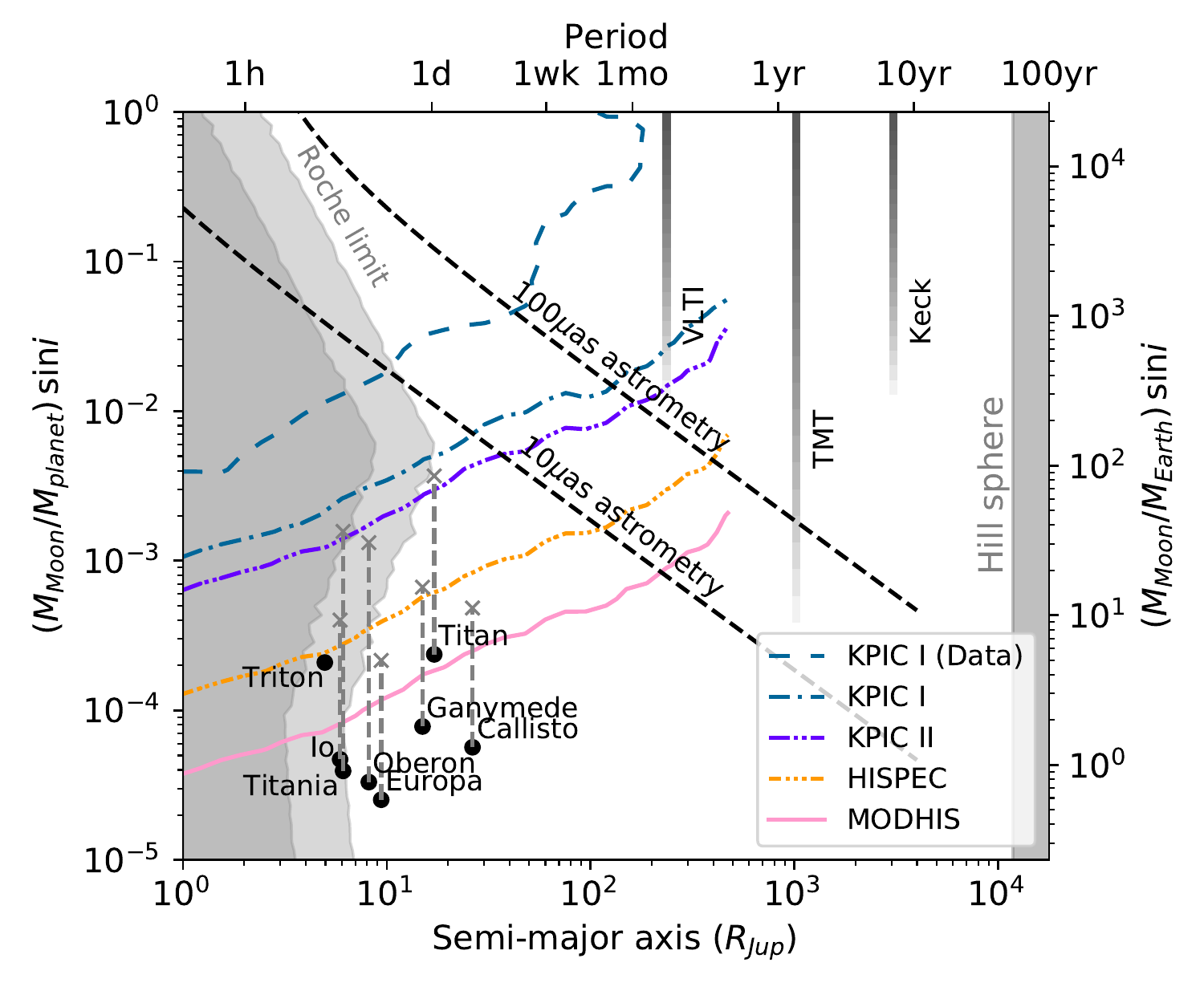}{0.49\textwidth}{(a) HR 7672 B}
		\fig{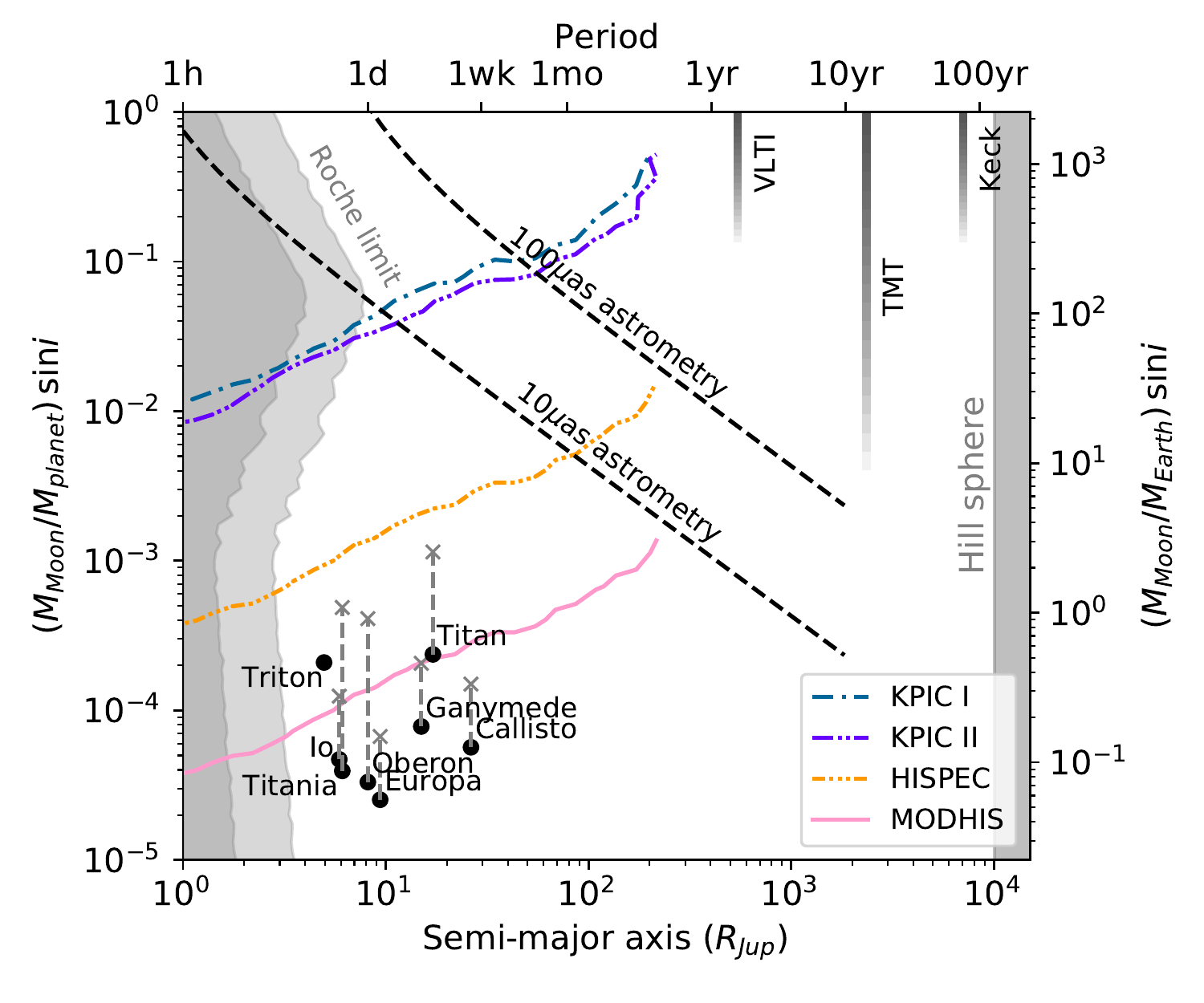}{0.49\textwidth}{(b) HR 8799 c}
          }
\caption{ Similar to \autoref{fig:moonprospects_RVonly}, but including idealized exomoon sensitivities of alternative detection techniques. The diagonal dashed black lines represent the simplified sensitivity of VLTI/Gravity through astrometry. The vertical gray scale bars represent the diffraction limit of different telescopes for direct imaging of satellites, namely the W. M. Keck observatory, the future Thirty Meter Telescope (TMT), and the Very Large Telescope Interferometer (VLTI).
}
\label{fig:moonprospects_comparison}
\end{figure*}

Alternative exomoon detection techniques include astrometry and direct imaging of imaged planets. \autoref{fig:moonprospects_comparison} shows their idealized detection limits to be compared to the RV sensitivity originally presented in \autoref{fig:moonprospects_RVonly}. 
With an astrometric precision of $10-100\,\mu\mathrm{as}$ \citep{Abuter2021}, interferometry with VLTI/GRAVITY could be sensitive to moons further away than radial velocity, but remains limited by the orbital period of the satellite at the furthest separations. The simplified detection limits are computed by matching the astrometric precision ($\sigma_{\mathrm{astro}}$) of VLTI/GRAVITY with the amplitude of the planet astrometric displacement in the sky around the center of mass. The smallest detectable mass ratio ($q$) is given by
\begin{equation}
q=\left[ 2*\left(\frac{\mathrm{sma}}{1\,\mathrm{au}}\right)\left(\frac{1\,\mathrm{pc}}{d}\right) \left(\frac{1\,\mathrm{as}}{\sigma_{\mathrm{astro}}}\right) -1 \right]^{-1},
\end{equation}
with $d$ the distance of the star to the Sun, and $\mathrm{sma}$ the semi major axis of the moon.
We use the diffraction limit of the telescope to illustrate the parameter space that might be accessible to direct imaging. More specifically, the detection threshold is taken at twice the spatial resolution of the telescope ($\sim2\lambda/D$) with $D$ the diameter of the telescope and $\lambda=2\,\mu\mathrm{m}$. Unfortunately, estimating the brightness of low mass objects ($<1M_{\mathrm{Jup}}$) remains challenging and will depend on the age of the system, so we arbitrarily chose a lower limit of one Jupiter mass for Keck and VLTI, and a mass similar to the solar system ice giants for TMT. Direct imaging would be sensitive to the longest periods and largest moons.

